\documentclass[11pt,a4paper]{article}

\usepackage[utf8]{inputenc}
\usepackage[T1]{fontenc}
\usepackage{lmodern}
\usepackage{amsmath, amssymb, mathtools, physics, cancel, xcolor}
\usepackage{graphicx}
\usepackage{setspace, parskip}
\usepackage{fancyhdr, caption, subcaption}
\usepackage{booktabs}
\usepackage{multirow}
\usepackage{float}
\usepackage[]{biblatex}
\usepackage{csquotes, tikz}

\usepackage[
    top=2.54cm, 
    bottom=2.54cm, 
    left=2.54cm, 
    right=2.54cm, 
    headheight=15pt, 
    headsep=1cm, 
    footskip=2cm
]{geometry}

\newcommand{\Ra}{\mathit{Ra}}
\newcommand{\Nu}{\mathit{Nu}}
\newcommand{\Pran}{\mathit{Pr}}

\usepackage{authblk}

\usepackage{hyperref}
\hypersetup{
    colorlinks = true,
    urlcolor   = blue,
    citecolor  = black,
}
\title{A Lagrangian analysis of coherent structures in 2D annular Rayleigh-Bénard convection}
\date{\small (Dated: \today)}
\author[1]{Luis Álamo\thanks{Corresponding author: luis.alamo@upc.edu}} 
\author[1,2]{Jezabel Curbelo}
\author[3]{Kathrin Padberg-Gehle}

\affil[1]{Departament de Matemàtiques, Universitat Politècnica de Catalunya, Barcelona 08028, Spain}
\affil[2]{Centre de Recerca Matemàtica, Campus de Bellaterra, Edifici C, Bellaterra 08193, Barcelona, Spain}
\affil[3]{Institute of Mathematics and its Didactics, Leuphana Universität Lüneburg, Lüneburg 21335, Germany}

\begin{document}
\maketitle

\begin{abstract}
Convective instabilities are studied from a Lagrangian perspective in order to gain new insights into how the resulting global, long-lived coherent flow structures organize themselves. We consider a 2D annular fluid layer uniformly heated from the inside and subject to radial gravity and propose several new Lagrangian diagnostics for characterizing its convective instabilities. In particular, we perform a comprehensive Lagrangian analysis of the flow field across a wide range of dimensionless parameters, establishing a heuristic correspondence between convective plumes and curves of local maxima of Finite-Time Lyapunov Exponent (FTLE) fields. Moreover, we characterize coherent swirling motions inside convective rolls, which are associated to the non-zero curvature of the flow. By introducing a sliding window method for our finite-time kinematic measures, we demonstrate a direct relationship with radial heat flux quantified by means of the Nusselt number and identify relevant dynamical transitions that leave a trace on both. Finally, we propose a formal criterion for determining viscous boundary layer (UBL) thickness by virtue of radial FTLE profiles and validated against a classical Eulerian method. 

\end{abstract} 	 

\section{Introduction}
\label{introduction}

Planar Rayleigh-Bénard Convection (PRBC), the standard model of thermal convection, is a benchmark in nonlinear dynamics and pattern formation. It consists of a buoyancy-driven fluid layer between two solid plates, where the bottom one is heated and the top one is cooled, see \cite{RBC_review_Lohse} for a review. Applications range from technology and engineering to geophysics and astrophysics, always associated to transport and mixing processes. In particular, thermal convection is the mechanism that triggers turbulent heat transport in many geophysical \cite{Vallis_Parker_Tobias_2019} and astrophysical \cite{sun_convection_review} flows. If the temperature gradient is large enough, a thermal instability occurs, and later on, a large scale circulation sets in. It is formed by long-term plumes and rolls, structures that have a fundamental impact on transport and mixing in the flow. In the Eulerian frame of reference, these large scale circulation patterns have been extensively studied in numerous different settings, all of them are motivated by particular problems or applications. This has {deepened our understanding of} how small changes in the parameters impact heat transport and mixing. However, many questions remain open, such as the physical mechanism that triggers the spontaneous halts and reversals of direction of this large scale circulation \cite{LSC_reversals}. From a Lagrangian perspective, since it is naturally associated to material transport, the large scale circulation is a coherent configuration, meaning that it keeps its spatial integrity without developing smaller scales. Lagrangian mathematical approaches allow us to quantitatively classify fluid subdomains that maximize (or minimize) heat transfer in the flow.

Numerous mathematical and computational methods based on concepts from dynamical systems theory have been proposed to study Lagrangian transport and mixing processes, resulting in different but related notions and definitions of coherent flow structures (see \cite{Allshouse2015,haller_LCS_review,Hadjighasem2017,Balasuriya2018, Badza2023,curbelo2025} for reviews and comparisons). Fundamental objects of interest in the present study are Lagrangian Coherent Structures (LCSs, \cite{haller_LCS_review,Haller_2023}). These are hypersurfaces that act as barriers to transport and mixing, reminiscent of invariant manifolds of hyperbolic objects in classical dynamical systems. LCSs are defined as material hypersurfaces that organize transport and deformation of fluid trajectories over a finite time interval. A variety of diagnostics and methods have been developed to identify such structures, including the Finite-Time Lyapunov Exponent (FTLE, \cite{SHADDEN2005,haller_LCS_review}), Lagrangian descriptors \cite{mancho2013} or the Lagrangian-Averaged Vorticity Deviation (LAVD, \cite{Haller_2016_LAVD}).  These approaches have been successfully applied to many different complex flows, including geophysical datasets \cite{Haller_Beron-Vera_2013, Haller_vortices_Pacific, curbelo2021, bruera2023, Curbelo2023, rivoire2026}. Geophysical applications have also been the motivation for other simple, heuristic measures based on trajectory interactions: studying close encounters of trajectories \cite{Rypina2017}, which can be quantified by means of the node degree in a Lagrangian trajectory network \cite{KPG2017} and formally related to FTLE \cite{KPG2019}.
 
Several Lagrangian analyses have been conducted in 2D and 3D PRBC \cite{KPG_RBC2021,KPG_RBC2022,VIEWEG202469}, and they have yielded new insights into the key role that LCSs have on vertical heat transport. Here, we investigate convection on a 2D annular domain in which the inner plate is hotter than the outer one and a radial inwardly increasing gravity is imposed. On the one hand, this setup is motivated by studying a simple model in order to assess the effect of geometry on the resulting convective coherent structures and, consequently, its impact on radial heat flux from a Lagrangian perspective. On the other hand, this 2D model has only recently been studied in the Eulerian frame of reference \cite{Bhadra_Shishkina_Zhu_2024, BHADRA2025126703} with the aim to understand relevant physical mechanisms in stratified convection, which may be one important driver of turbulence in cold regions inside protoplanetary disks (rotating dusty regions around young newly formed stars). The present work is to the best of the authors' knowledge the first Lagrangian study of this underexplored configuration.
 
We also aim to find relations between the kinematics of mixing from the Lagrangian perspective and radial heat flux, so the question we address in this paper is the following: to what extent does the Lagrangian approach add to the classical Eulerian convection analyses? From this, central objects of the present study are boundary layers (BLs). In particular, we propose a new method to identify viscous boundary layers (UBLs) limits by using azimuthally-averaged FTLE profiles. As passive tracers get stuck inside UBLs, we expect a significant kinematic transition when they detach and merge into the base flow. 

The remainder of this paper is organized as follows. In section \ref{sec:methods}, we briefly review the basic aspects of the different Lagrangian techniques we implement, from the geometrical tools {based on FTLE and LAVD} to network-based ones. We also introduce the mathematical annular RBC model and provide a detailed explanation of the computational steps needed to compute all Lagrangian analysis quantities. Then, in section \ref{sec:result}, we present our results: a complete Lagrangian description of the flow field for different parameter values, a new Lagrangian method for identifying relevant dynamical transitions that allow us to compare with an instantaneous measure of the Nusselt number, and finally, a detailed explanation of the previously mentioned Lagrangian approach to UBL thickness detection. Section \ref{sec:discussion} is devoted to a discussion and an outlook of the results.

\section{Methods}
\label{sec:methods}

\subsection{Basic aspects of the Lagrangian approaches in 2D}
\label{sec:theory}
In this section, we briefly introduce the different Lagrangian techniques that will be used in our study. Thus, given a fluid defined over the physical domain $\Omega_t$ at each $t \in I$, the movement of a single passive tracer at $\mathbf{x}_t$ is completely determined by the initial value problem
\begin{align}\label{eq:campo_velo}
        & \left\{ \begin{aligned}
        &\dot{\mathbf{x}}_t=\mathbf{u}(t, \mathbf{x}_t), \\
        &\mathbf{x}_{t_0} = \mathbf{x}_0,
        \end{aligned}\right. 
    \end{align}
where the velocity field $\mathbf{u}  : \mathcal{U}\subset I \times \mathbb{R}^2 \rightarrow \mathbb{R}^2$ is a function defined on the open and connected subset $\mathcal{U}=\{(t, \mathbf{x})\in I \times \mathbb{R}^2 : t\in I, \: \mathbf{x}\in\Omega_t\}$. If we fix $I=[t_0, t]$ and take together all the initial positions of the fluid at some time $t_0$, that is to say, all the $\mathbf{x}_0 \in \Omega_{t_0}$, the general solution of \eqref{eq:campo_velo} is called \textit{flow} and is given by
\begin{equation}\label{eq:deformaciones}
        \mathbf{F}_{t_0}^t : \mathbf{x}_0 \in \Omega_{t_0} \mapsto \mathbf{F}_{t_0}^t(t; t_0, \mathbf{x}_0) = \mathbf{x}_t\in \Omega_t.
    \end{equation}
Assuming enough regularity of $\mathbf{u}$, the flow map is a diffeomorphism. Thus, taking derivatives with respect to the initial positions $\mathbf{x}_0$ we consider the deformation gradient $\nabla\mathbf{F}_{t_0}^t(\mathbf{x}_0)$ such that
\begin{equation}
        \nabla\mathbf{F}_{t_0}^t(\mathbf{x}_0) :  T_{\mathbf{x}_0} \mathbb{R}^2 \longrightarrow T_{\mathbf{F}_{t_0}^t(\mathbf{x}_0)}\mathbb{R}^2.
    \end{equation}
Physically, $\nabla\mathbf{F}_{t_0}^t(\mathbf{x}_0)$ is a tensor that describes the evolution of an infinitesimal deformation $\xi_0$ of $\mathbf{x}_0$ to a vector $\nabla\mathbf{F}_{t_0}^t(\mathbf{x}_0)\xi_0$ for some time $t$. 

We define the (right) Cauchy-Green strain tensor $\mathbf{C}_{t_0}^t(\mathbf{x}_0) := (\nabla\mathbf{F}_{t_0}^t(\mathbf{x}_0))^\top \nabla\mathbf{F}_{t_0}^t(\mathbf{x}_0)$, which is symmetric and positive definite (here $\top$ denotes transposition). Its largest eigenvalue $\lambda_{\text{max}}(t; t_0, \mathbf{x}_0)$, which is an objective quantity see \cite{Haller_2023}, is a measure of the growth of the infinitesimal perturbation. From this, we define the Finite-Time Lyapunov Exponent (FTLE) as the objective scalar field
\begin{equation}\label{eq:FTLE}
        \text{FTLE}_{t_0}^{t}(\mathbf{x}_0) = \frac{1}{2|t-t_0|} \log{\lambda_{\text{max}}(t; t_0, \mathbf{x}_0)}.
    \end{equation}    

Thus, we will refer to \textit{forward} (\textit{backward}) FTLE, or \textit{f}-FTLE (\textit{b}-FTLE) when $t > t_0$ ($t < t_0$). Hyperbolic LCSs will generically align with ridges of $\text{FTLE}_{t_0}^{t}(\mathbf{x}_0)$. In particular, repelling LCSs will generically align with ridges of the \textit{f}-FTLE and attracting LCSs with ridges of the \textit{b}-FTLE see \cite{Haller_2023}. As we are considering the 2D case, our LCSs will be just material curves.

As we expect our flow to be driven by convective cells that rotate around hypothetical vortex centers, we are also interested in assessing rotational coherence. Thus, we will use the Dynamic Polar Decomposition (DPD) of $\nabla\mathbf{F}_{t_0}^t(\mathbf{x}_0)$ developed by \cite{Haller_2016_DPD}. The main element of this decomposition is the relative rotation tensor $\Phi_{t_0}^t$, which is dynamically consistent. {The} intrinsic dynamic rotation angle $\psi_{t_0}^t(\mathbf{x}_{0})$ around the axis of rotation $-\frac{\omega-\overline{\omega}}{|\omega-\overline{\omega}|}$ is defined as the total accumulated rotation angle generated by $\Phi_{t_0}^t$, i.e.,
\begin{equation}
        \psi_{t_0}^t (\mathbf{x}_{0}) = \frac{1}{2}\int_{t_0}^t |\omega(s, \mathbf{x}_s)-\overline{\omega}(s)| \: \mathrm{d}s.
    \end{equation}
Note that this angle is computed around the axis defined by the local relative vorticity vector $\omega(s, \mathbf{x}_s)-\overline{\omega}(s)$, where $\overline{\omega}(s)$ denotes a spatial average over $\Omega_s$. Moreover, this objectively defined purely mathematical quantity is then related to a non-objective physical quantity: the vorticity. The Lagrangian-Averaged Vorticity Deviation, LAVD see \cite{Haller_2016_LAVD} is defined as the Lagrangian scalar field 
\begin{equation}\label{eq:LAVD}
        \text{LAVD}_{t_0}^t(\mathbf{x}_0) = \frac{2\psi_{t_0}^t(\mathbf{x}_0)}{t-t_0},
    \end{equation}
that will immediately inherit the dynamical consistency and objectivity of $\psi_{t_0}^t(\mathbf{x}_{0})$. From this, we refer to elliptic LCSs as smooth, closed and convex level curves of $\text{LAVD}_{t_0}^t(\mathbf{x}_0)$ surrounding a unique local maximum. Thus, a Lagrangian vortex will be a nested family of elliptic LCSs. 

As a complementary approach, we briefly describe the construction of the trajectory-based weighted network presented by \cite{KPG_RBC2022}. For this, let us consider $N$ tracer trajectories $\mathbf{x}_{i,t}$ ($i=1, \dots, N$) evaluated at discrete time slices of the spatial domain $\Omega_t$, with $t \in \mathbb{T}=\{t_0, t_0 + 1, \dots, t_0 + T\}$. These trajectories are solutions of \eqref{eq:campo_velo}, i.e.,
\begin{align}\label{eq:discrete_dyn_sys}
        & \left\{ \begin{aligned}
        &\dot{\mathbf{x}}_{i,t} = \mathbf{u}(t, \mathbf{x}_{i,t}),\\
        &\mathbf{x}_{i,t_0} = \mathbf{x}_{i,0}, 
        \end{aligned}\right. \quad \text{for all} \: i = 1,\dots , N.
\end{align}
We also fix $\varepsilon > 0$ and define an instantaneous adjacency matrix $A_t \in \{0, 1\}^{N\times N}$ such that a link between $\mathbf{x}_{i,t}$ and $\mathbf{x}_{j,t}$ ($i\neq j$) is established at a fixed time $t$ if, and only if, both tracers become $\varepsilon$-close at $t$. In other words,
\begin{align}\label{eq:adjacency_matrix}
        & A_{t, ij} = \left\{ \begin{aligned}
        &1, \quad \text{if} \quad \lVert \mathbf{x}_{i,t} - \mathbf{x}_{j,t}\rVert < \varepsilon \: (i\neq j)\\
        &0, \quad \text{otherwise}.
        \end{aligned}\right. 
    \end{align}
This network depends on a suitable definition of its unique free parameter $\varepsilon$, which should be chosen such that the network is connected. In our particular case of study, we shall compute distances in a non-uniform grid with nodes accumulating near its boundary (details in section \ref{subsec:sim_setup}), so an appropriate choice of $\varepsilon$ becomes difficult. To overcome this, we use a mutual $K$ nearest neighbors construction  \cite{VIEWEG202469}. Thus, a link between $\mathbf{x}_{i,t}$ and $\mathbf{x}_{j,t}$ is established if, and only if, they are $K$ mutual nearest neighbors, meaning that $\mathbf{x}_{i,t}$ belongs to the $K$ nearest neighbors of $\mathbf{x}_{j,t}$ and vice versa. The choice of $K$ allows us to control particle density and network connectivity, especially in boundary layer regions where particles become sticky for long times.

Let $W := \sum_{t\in\mathbb{T}} A_t$ be the network weight matrix, whose component $W_{ij}$ encodes the number of encounters of tracers $\mathbf{x}_{i,t}$ and $\mathbf{x}_{j,t}$ over the whole time span. Thus, $W_{ij}$ will be large when $\mathbf{x}_{i,t}$ and $\mathbf{x}_{j,t}$ have been close for a long time, in the sense of being mutual neighbors for many time instances. From $W$ we can define a measure of the total number of encounters that $\mathbf{x}_{i,t}$ has with any other tracer of the weighted network over time, that is $d_{w,i} = \sum_{j=1}^N W_{ij}$. This is often referred as \textit{weighted} node degree. 
Let $\overline{W} \in \{0,1\}^{N\times N}$ denote the adjacency matrix of the network, where $\overline{W}_{ij} =1$  if $W_{ij} >0$ and $=0$ otherwise. We can define an \textit{unweighted} node degree such that $d_{u,i} = \sum_{j=1}^N \overline{W}_{ij}$, which encodes how many \emph{different} tracers $\mathbf{x}_{i,t}$ has encountered over the whole time span.

Of particular interest is the number of different links established by $\mathbf{x}_{i,t}$ in relation to the total links weights, which is
\begin{equation}
    d_i = \frac{d_{u,i}}{d_{w,i}},
\end{equation}
the \textit{normalized} node degree. Just like the FTLE, we expect this measure to be large in highly mixing regions, where every tracer is encountering many different tracers over time. In fact, an analytical relation between the unweighted node degree and the FTLE was established by \cite{KPG2019}.

Finally, we note that $W$ can be used to identify coherent sets as regions in space with different dynamical behavior. These correspond to nearly isolated subgraphs in the network and can be computed by spectral clustering approaches \cite{KPG_RBC2022}. However, this is out of the scope of this article.

\begin{figure}
	\centering 
	\includegraphics[width=0.5\textwidth]{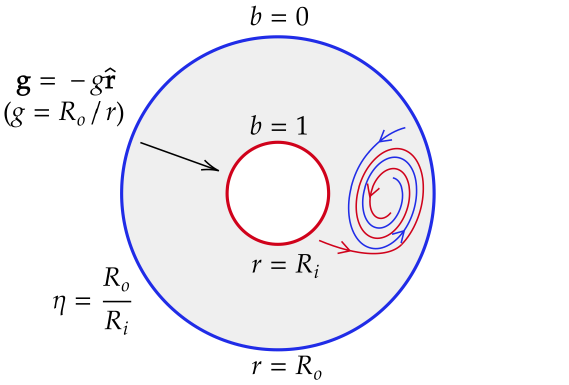}	
	\caption{Schematic diagram of the numerical setup. We show high temperatures in red and low temperatures in blue. An illustration of a single convective cell, formed by a hot and a cold plume, is shown.} 
	\label{fig:config}%
\end{figure}

\subsection{Physical setup}
\label{subsec:sim_setup}
The physical setup for this study corresponds to the 2D incompressible Rayleigh-Bénard convection equations in an annular fluid layer that is heated from the inside and cooled from the outside under the Boussinesq approximation. We have an inward gravity contribution $\mathbf{g}=-g\hat{\mathbf{r}}$ with $g \equiv R_o/r$, where $\hat{\mathbf{r}}$ denotes the unit vector in the radial direction, $R_o$ denotes the outer radius and $R_i$ the inner radius, as shown in figure \ref{fig:config}. We thus have a spatial domain $\Omega := \{(r, \phi)\in \mathbb{R}^2 \: : \: r\in [R_i, R_o], \: \phi\in [0, 2\pi]\} \subset \mathbb{R}^2$. In this way, gravity provides buoyancy and, therefore, drives the dynamics of the system. The relevant geometrical parameters are summarized by a radius ratio $\eta= R_o/R_i$, such that $R_i = 2/(1+\eta)$ and $R_o = 2\eta / (1+\eta)$. Thus, dimensionless equations for the buoyancy $b$ and the velocity $\mathbf{u}$ are
\begin{align}\label{eq:RB_Boussinesq_adim_simu}
        \displaystyle\nabla\cdot\mathbf{u} & = 0, \\
        \frac{\partial \mathbf{u}}{\partial t} + (\mathbf{u} \cdot\nabla)\mathbf{u} & = -\nabla p + \sqrt{\frac{\Pran}{\Ra}} \Delta\mathbf{u} + b\hat{\mathbf{r}}, \\
        \displaystyle\frac{\partial b}{\partial t} + (\mathbf{u}\cdot\nabla)b & = \frac{1}{\sqrt{\Pran \: \Ra}}\nabla^2 b,
    \end{align}
where $p$, $b$ and $t$ denote the dynamic pressure, the buoyancy (or just dimensionless temperature) and time, respectively. In components, these read as
\begin{align}\label{eq:RB_Boussinesq_adim_simu_comp}
        & \displaystyle\frac{\partial(r u_r)}{\partial r} + \frac{\partial u_\phi}{\partial\phi} = 0, \\
        \begin{split}
        & \displaystyle\frac{\partial u_r}{\partial t} + u_r \frac{\partial u_r}{\partial r} + \frac{u_\phi}{r}\frac{\partial u_r}{\partial r} - \frac{u_\phi^2}{r} = -\frac{\partial p}{\partial r}\\
        & \displaystyle + \sqrt{\frac{\Pran}{\Ra}} \left[\frac{1}{r}\frac{\partial}{\partial r}\left(r\frac{\partial u_r}{\partial r}\right) + \frac{1}{r^2}\frac{\partial^2 u_r}{\partial\phi^2} - \frac{u_r}{r^2} - \frac{2}{r^2}\frac{\partial u_\phi}{\partial \phi}\right] + bg,
        \end{split}\\
        \begin{split}
        & \displaystyle\frac{\partial u_\phi}{\partial t} + u_r \frac{\partial u_\phi}{\partial r} + \frac{u_\phi}{r}\frac{\partial u_\phi}{\partial \phi} + \frac{u_\phi u_r}{r} = -\frac{1}{r}\frac{\partial p}{\partial \phi}\\
        & \displaystyle + \sqrt{\frac{\Pran}{\Ra}} \left[\frac{1}{r}\frac{\partial}{\partial r}\left(r\frac{\partial u_\phi}{\partial r}\right) + \frac{1}{r^2}\frac{\partial^2 u_\phi}{\partial\phi^2} - \frac{u_\phi}{r^2} + \frac{2}{r^2}\frac{\partial u_r}{\partial \phi}\right],
        \end{split}\\
        & \displaystyle\frac{\partial b}{\partial t} + u_r\frac{\partial b}{\partial r} + \frac{u_\phi}{r}\frac{\partial b}{\partial \phi} = \frac{1}{\sqrt{\Pran \: \Ra}} \left[\frac{1}{r}\frac{\partial}{\partial r}\left(r \frac{\partial b}{\partial r}\right) + \frac{1}{r^2}\frac{\partial^2 b}{\partial \phi^2}\right],
    \end{align}
where $u_r$, $u_\phi$ denote the radial and angular components of the velocity field, i.e., $\mathbf{u}=(u_r, u_\phi)$. The parameters involved in the non-dimensionalizing process are the mean radius $d=(R_o+R_i)/2$, the difference in temperature between plates $\Delta b = b(r=R_i)-b(r=R_o)$ and the free-fall velocity $U_c=\sqrt{\alpha g_o d \Delta b}$, where $g_o$ denotes the gravitational constant at $r=R_o$ and $\alpha$ denotes the isobaric expansion coefficient. As a consequence, time is measured in free-fall time units ($T_c$) such that
\begin{equation}
    T_c = \frac{d}{U_c} = \sqrt{\frac{d}{\alpha g_o \Delta b}}.
\end{equation}

Moreover, the control parameters derived from these are the Rayleigh number $\Ra = \alpha g_o d^3 \Delta b / (\nu\kappa)$, the Prandtl number $\Pran=\nu/\kappa$ and the previously defined radius ratio $\eta$. In this context,  $\nu$ is the kinematic viscosity and $\kappa$ the thermal diffusivity. They are computed in terms of the dimensionless numbers such that
\begin{equation}
    \kappa = \frac{1}{\sqrt{\Ra \: \Pran}}, \quad \nu = \sqrt{\frac{\Pran}{\Ra}}.
\end{equation}
In all our simulations, we fix $\Pran=1$ and we also use no-slip and isothermal boundary conditions.

In particular, we perform simulations with four different radius ratios $\eta\in\{1.25, 1.5, 2.5, 5\}$ and $\Ra$ ranging from $5 \cdot 10^4$ to $1 \cdot 10^7$. This selection of parameters is motivated by covering $\Ra$ near its critical value and the fact that coherent structures vanish rapidly for large $\Ra$, as noted by \cite{Bhadra_Shishkina_Zhu_2024} from their Eulerian point of view. In fact, this range of $\Ra$ is considered laminar, as the corresponding bulk flow is not \textit{fully} turbulent. It also makes it more likely to obtain interesting dynamical behaviors in a controlled environment, such as splitting of plumes. This will be relevant in section \ref{sec:heat_flux}.

We provide our simulations with a conductive background profile given by
\begin{equation}\label{eq:cond_profile}
    b_c = \log(r/R_o) / \log(\eta^{-1}),
\end{equation}
thus $\mathrm{d}b_c/\mathrm{d}r = (r\log(\eta^{-1}))^{-1}$. This profile is motivated by the analytical solution reported by \cite{Ali_Weidman_1990} (and used, for example, by \cite{Tagg_Weidman_1990}) for the steady flow in radially-heated concentric and rotating cylinders. This also allows us to make some useful comparisons with \cite{Bhadra_Shishkina_Zhu_2024, BHADRA2025126703}. The total flux in the radial direction is given by $\Phi_{r, \text{tot}} = -\kappa \: \mathrm{d}b / \mathrm{d}r + u_r b$. Therefore, the Nusselt number (in the Eulerian reference frame) is computed such that
\begin{equation}
    \displaystyle \Nu = \frac{\langle \Phi_{r, \text{tot}} \rangle_{\Omega , t}}{\Phi_{r, \text{c}}} = \frac{\langle u_r b\rangle_{\Omega , t} - \frac{1}{\sqrt{\displaystyle \Ra \: \Pran}} \frac{\mathrm{d}\langle b \rangle_{\Omega , t}}{\mathrm{d}r}}{\frac{-1}{\sqrt{\displaystyle \Ra \: \Pran}} \frac{\mathrm{d}b_c}{\mathrm{d}r}},
\end{equation}
where $\langle \: \cdot \: \rangle_{\Omega , t}$ denotes a combined average in time $t$ and the whole spatial domain $\Omega$. Throughout this study, we will also use other averages of the Nusselt number with different purposes. This will be specified when necessary. Furthermore, we compute the root mean square of the velocity as
\begin{equation}
    u_{rms} = \sqrt{\langle u_\phi^2 \rangle_{\phi, t} + \langle u_r^2 \rangle_{\phi, t}}.
\end{equation}

Finally, the numerical method implemented in \texttt{Dedalus3} consists of a spectral \textit{tau} method with two \textit{tau} terms for the velocity and two more for the buoyancy (matching the boundary conditions). We also need to fix the pressure gauge for non-degeneracy of the incompressibility condition, as noted by \cite{Dedalus_article}. The spectral basis that we use is a direct product of a Fourier basis in the angular direction with $n_\phi$ nodes and a Chebyshev basis in the radial direction with $n_r$ nodes given by a Gauss-Chebyshev quadrature. Table \ref{table:datas} shows a summary of the parameters involved in all simulations performed. For the temporal evolution, we use the \texttt{RK443} timestepper implemented in Dedalus, which is a third-order four-stage Runge-Kutta implicit-explicit scheme.

\subsection{Computation of Lagrangian features}
For all three methods described in section \ref{sec:methods}, Lagrangian trajectories need to be computed from the simulated velocity field (i.e., the solutions of \eqref{eq:campo_velo}) and this requires an interpolation routine at every step of a Runge-Kutta scheme. To this end, we make use of the \texttt{Python} library \texttt{NumbaCS} \cite{Jarvis2025}, where a series of packages are combined to optimize the numerical procedures. First, we create a cubic spline over an auxiliary uniform grid in which we evaluate the velocity. We then obtain a whole set of finite-time Lagrangian trajectories (i.e., the flow \eqref{eq:deformaciones}) by integrating the interpolated velocities. Note that, as velocities are computed in polar coordinates, the ODE to be solved takes the form
\begin{align}\label{eq:campo_velo_polar}
        & \left. \begin{aligned}
        \displaystyle \dot{r}_t & =u_r(t, r_t), \\
        \displaystyle \dot{\phi}_t & =\frac{u_\phi(t, \phi_t)}{r}.
        \end{aligned}\right. 
    \end{align}
We use \texttt{dop853} as ODE solver, which is an explicit Runge-Kutta method of order 8. {Trajectories} are computed from data in polar coordinates, so units are arbitrary both in the radial and angular directions. 
From these trajectories, we obtain the Lagrangian fields of interest as follows:

\begin{enumerate}
    \item We compute the deformation gradient under a coordinate transformation from polar to cartesian, such that
        \begin{equation}
            \nabla\mathbf{F}_{t_0}^t = \frac{\partial (x,y)}{\partial (x_0, y_0)} = \frac{\partial (x,y)}{\partial (r_0, \phi_0)} \frac{\partial (r_0,\phi_0)}{\partial (x_0, y_0)},
        \end{equation}
    where $x$ and $y$ denote the actual cartesian coordinates obtained from $r$ and $\phi$ and $x_0$, $y_0$, $r_0$ and $\phi_0$ denote the coordinates of the initial position in both cartesian and polar reference frames. Thus, the notation $\partial (\cdot,\cdot) / \partial (\cdot, \cdot)$ is interpreted as the Jacobian of the corresponding transformation. In this way, we apply a finite differences scheme of second order to compute $\partial (x,y) / \partial (r_0, \phi_0)$ and we explicitly compute
    \begin{equation}
		\frac{\partial (r_0,\phi_0)}{\partial (x_0, y_0)} = \begin{pmatrix}
			\displaystyle \cos{\phi_0} & \displaystyle \sin{\phi_0}  \\
			\displaystyle\frac{-\sin{\phi_0}}{r_0} & \displaystyle\frac{\cos{\phi_0}}{r_0} \\
		\end{pmatrix}.
	\end{equation}
    Finally, we compute a finite-differences approximation of the Cauchy-Green strain tensor and look for its maximum eigenvalue. Equation \eqref{eq:FTLE} allows us to compute the FTLE field. 

    \item To compute the LAVD field, we build another spline defined over the same uniform grid as in the case above. We compute the (scalar) 2D vorticity in polar coordinates over it such that
    \begin{equation}
        \omega := \text{rot}_{2D} (\mathbf{u}) = \frac{1}{r} \left(\frac{\partial}{\partial r}(r u_r) - \frac{\partial u_r}{\partial \phi}\right), 
    \end{equation}
    where $\text{rot}_{2D}$ denotes the third component of the 3D rotational operator. From this, we build a spline of the 2D vorticity that is given as an argument of our LAVD function. We then compute the LAVD as defined in equation \eqref{eq:LAVD}.
    \item For the network-based approach, we can directly work with the given tracer trajectories, without the need of additional interpolation. The distances between tracers required for identifying $K$ mutual neighbors in the network construction are computed in a cartesian grid. The neighborhood search is efficiently carried out applying MATLAB's \texttt{knnsearch}, which uses a $K$d-tree. 
\end{enumerate}

The definition of an insightful time interval of interest is a crucial aspect of these finite-time Lagrangian-based approaches. Its choice will determine entirely the dynamical system defined by \eqref{eq:campo_velo} and will also affect the trajectories computed from the simulated velocity field. 
In other words, we need to select an initial time $t_0$ and some time $t_0 \pm T$ with $T$ being the integration time used for the computation of trajectories (both forward and/or backward). In this article, $t_0$ and $T$ will change depending on the main purpose of the corresponding section, but it will always be coherent with the underlying dynamics of interest. This will be specified when necessary.

\subsection{Sliding window approach}
\label{subsec:sliding_window}
In section \ref{sec:heat_flux}, we will study the dynamics of 2D coherent structures over long time spans in the Lagrangian reference frame. In particular, we seek to observe a relation between traces of significant changes in their dynamical behavior (such as mergers or splits of plumes), the radial heat flux (or Nusselt number) and the quantities defined in section \ref{sec:theory}. In order to do this, we propose a sliding window approach in the same spirit as the evolutionary network method proposed by \cite{KPG_RBC2022} both for the normalized node degree and the \textit{f}-FTLE. 

In the normalized node degree case, we start from forward and backward trajectories computed from $t_0$ to $t_0 \pm T/2$. Then, we compute weight matrices $W_{t_0}$ over these windows of length $T$ centered at $t_0$, so that $\mathbb{T}_{t_0, T} = \{t_0-T/2, \dots, t_0+T/2\}$ and $W_{t_0} = \sum_{s\in\mathbb{T}_{t_0, T}} A_s$. This construction allows us to carry less material information for longer time spans and, if $T$ is small enough, we can expect $W_{t_0}$ to vary smoothly with time.

For the \textit{f}-FTLE we just compute forward trajectories and the corresponding \textit{f}-FTLE field on time windows of length $T$ and we slide $t_0$ in order to cover long time spans.

\begin{figure}
	\centering 	\includegraphics{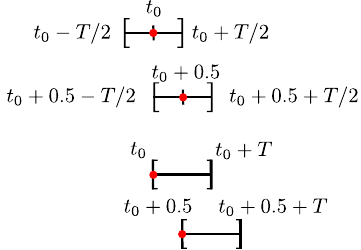}	
	\caption{Two-step diagram of the sliding window approach used in section \ref{sec:heat_flux} for both normalized node degree (above) and \textit{f}-FTLE (below). Red points indicate the exact points where we are plotting each technique. See section \ref{subsec:sliding_window} for more details.} 
	\label{fig:sliding_window}%
\end{figure}

To this end, we always set $T=1$ to compute a spatial average of the normalized node degree at the center of each integration time window and a spatial average of the \textit{f}-FTLE. Thus, we have two spatially averaged quantities such that

\begin{align}   \langle\text{FTLE}_{t_0}^{t_0+T}\rangle_{\Omega} & = \frac{1}{\text{vol} \: \Omega} \int_0^{2\pi} \int_{R_i}^{R_o} \text{FTLE}_{t_0}^{t_0+T}(\mathbf{x}_0) \: \mathrm{d}\mathbf{x}_0,\label{eq:FTLE_mean}\\
    \langle d_{t_0}^{t_0\pm T/2} \rangle_{\Omega} & = \frac{1}{N}  \sum_{j=1}^N d_{t_0,j}^{t_0\pm T/2}, \label{eq:node_mean}
\end{align}
where $\text{vol}\: \Omega$ denotes the volume of the spatial domain and $d_{t_0,j}^{t_0\pm T/2}$ the node degree of tracer $j=1, \ldots, N$, computed over $[t_0-T/2, t_0+T/2]$. We always slide $t_0$ in steps of 0.5 time units. This choice reflects the dynamics we are looking for quite well. As a consequence, we are able to plot \textit{smooth} spatially averaged time curves of the normalized node degree and the \textit{f}-FTLE over long time spans, in such a way that both curves are plotted at the same time \textit{nodes}. A schematic diagram of this method is shown in figure \ref{fig:sliding_window}.

\section{Results}
\label{sec:result}

In this section, we outline the results obtained by applying the Lagrangian approaches to the simulated convective flow. It is structured as follows: First, we present an exploratory study of the fluid flow and discuss the outcome of applying the techniques described in section \ref{sec:theory}. Figure \ref{fig:fig3} shows all the geometrical configurations for a fixed $\Ra = 10^6$, and figure \ref{fig:swirl_advected} illustrates the dynamics inside a convective cell for a particular case with $\Ra = 10^7$. Then, we discuss the relation between these kinematic measures of mixing with radial heat flux, illustrated in figure \ref{fig:Nu_fftle} for a fixed $\Ra = 10^6$. From this, we explore some insightful dynamics in figures \ref{fig:fig6} and \ref{fig:Nu25_1e6} for different $\eta$. Finally, we address boundary layer detection in section \ref{sec:TBLs}. We show some examples for a fixed $\Ra = 10^6$ in figure \ref{fig:UBLs}, all buoyancy radial profiles in figure \ref{fig:buoy_profiles}, our viscous boundary layer (UBL) estimates for a particular case with $\Ra = 10^7$ in figure \ref{fig:UBLS_T} and a summary of UBL results for all $\Ra$ in table \ref{table:datas}. 

\subsection{Lagrangian analysis of the flow field}
\label{sec:flow_field}

\begin{figure*}
        \centering
        \includegraphics[width=0.95\textwidth]{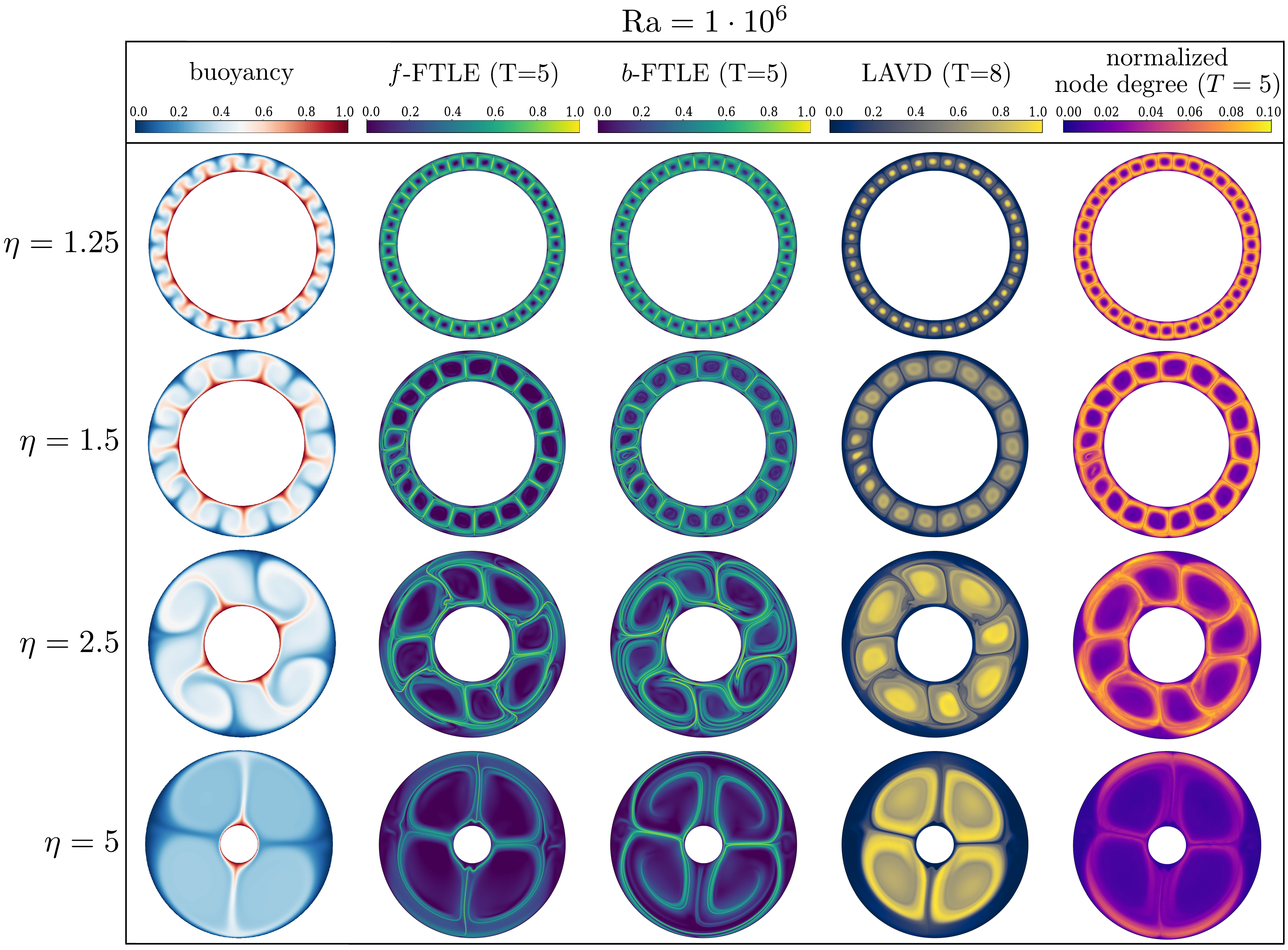}
    \caption{Summary table of all Lagrangian techniques and all annular widths ($\eta$) considered for a fixed Rayleigh number. All color maps are normalized for plotting purposes.}
    \label{fig:fig3}%
\end{figure*}

In all our simulations we begin with an initial equilibrium solution given by \eqref{eq:cond_profile} that, when perturbed, will evolve until thermal instability occurs. Then, the flow dynamics are dominated by a number of convective cells that depends on $\eta$ ---the thinner the annulus, the more cells---, each one driven by a hot inner plume that rises and a cold outer plume that descends. This is exactly the hyperbolic skeleton of transport we detect with the FTLE, that is to say, we find both attracting and repelling LCSs delineating each convective cell (see figure \ref{fig:fig3}). 

In particular, hot plumes correspond to the regions of largest \textit{f}-FTLE values, while cold plumes correspond to regions of the largest \textit{b}-FTLE values thus, as buoyancy is our driving force, we can heuristically relate hot plumes to repelling LCSs and, correspondingly, cold plumes to attracting LCSs. However, as this is a 2D incompressible flow, we cannot assure that every \textit{f}-FTLE (\textit{b}-FTLE) ridge (or curve of maximum values) is repelling (attracting) see \cite{karrasch2015}. We always need some additional postprocessing, such as some physical properties of the fluid flow.

Moreover, these flow structures are a consequence of mass conservation (or area-preserving flow), i.e., any outward mass flow must be compensated by a corresponding inward flow. But, unlike in the planar Rayleigh-Bénard problem, convective cells are distorted near both rigid plates by the annular geometry of the domain, that is, convective cells become wider near the outer plate and thinner near the inner plate. Radial velocities in the hot plumes exceed those in the cold plumes \cite{Bhadra_Shishkina_Zhu_2024}, resulting in a shear stress between the convective cell and the outer plume. In extreme cases (large $\Ra$ and $\eta$), these outer plumes tend to vanish before reaching the inner plate.

\begin{figure}
        \centering
        \includegraphics[width=0.7\textwidth]{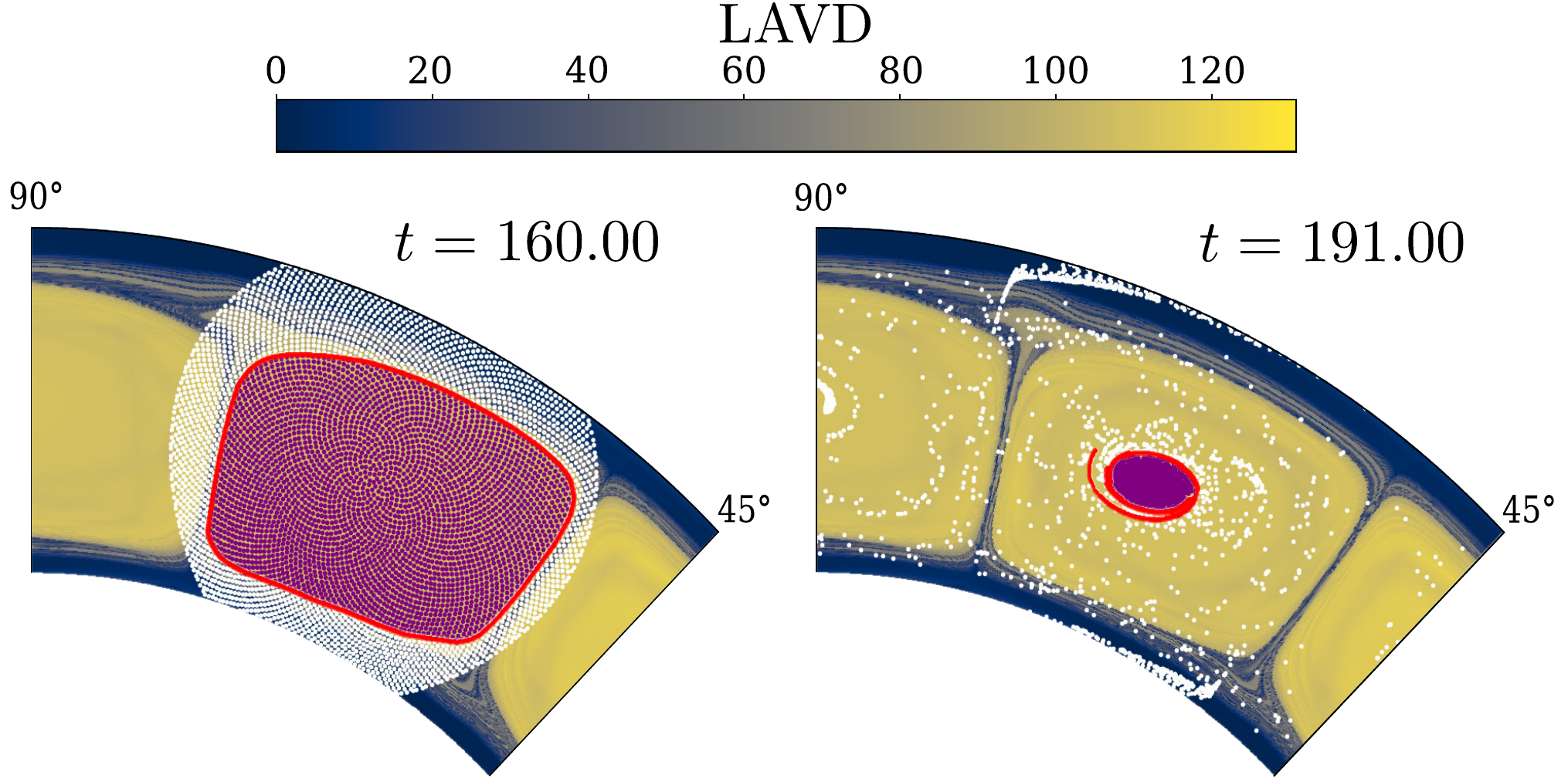}
        \raisebox{0.5cm}{\includegraphics[width=0.35\textwidth]{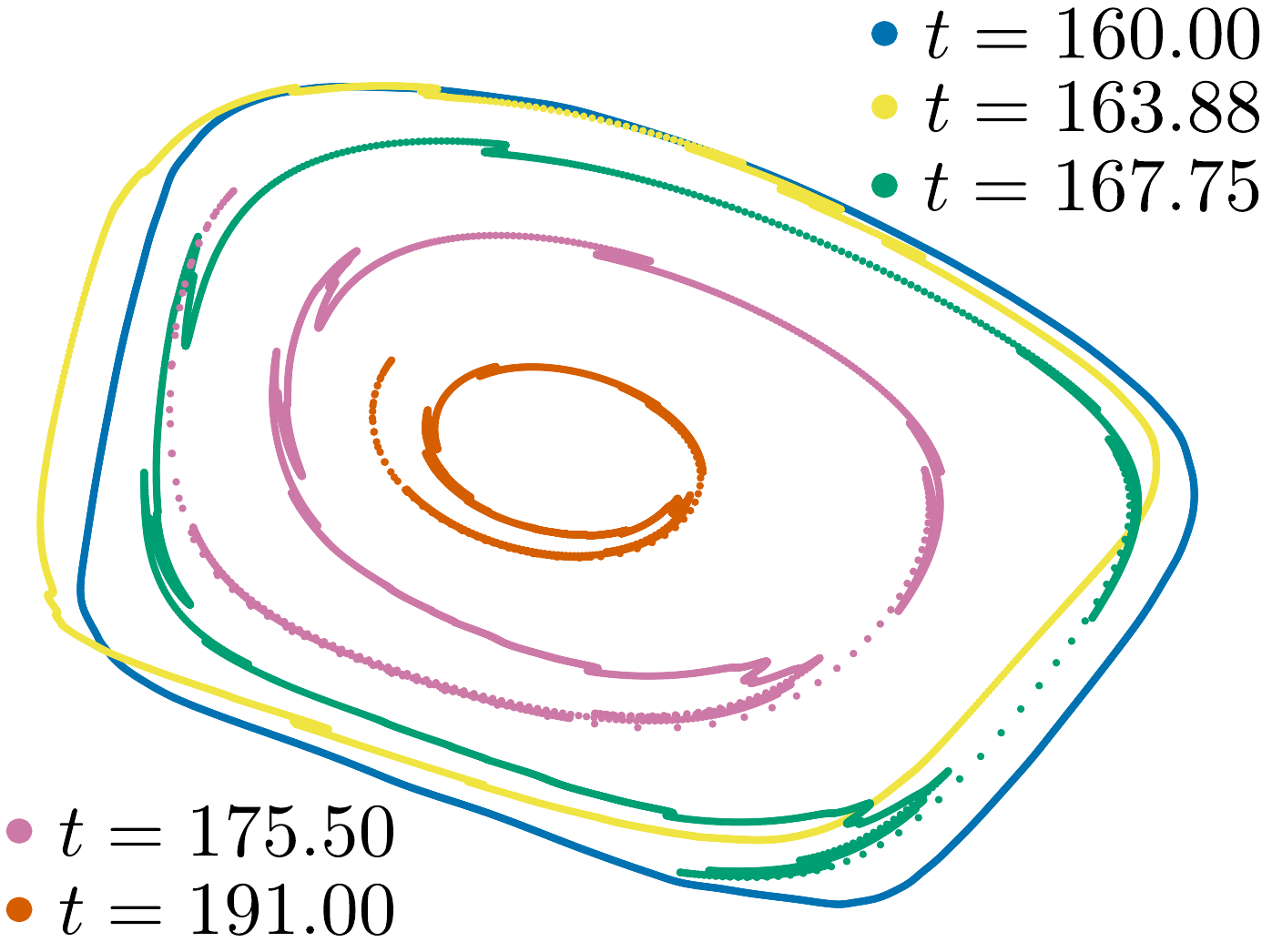}}
        \includegraphics{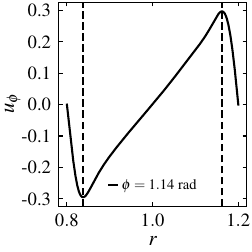}
    \caption{Above: initial (left) and final (right) animation frames of flow advected tracers around a detected Coherent Lagrangian Swirl inside a convective cell. Below: time evolution of the CLS boundary tracers over some selected moments in time (left). We can clearly see by the density of points where stretching and folding are maximized. Angular velocity profile at the exact detected $\phi$-coordinate of the CLS center (right). This example corresponds to $\Ra=10^7$, $\eta=1.5$.}
    \label{fig:swirl_advected}%
\end{figure}

Comparing these structures with the ones obtained using the network-based normalized node degree, we see that this technique properly summarizes the forward and backward FTLE information, as expected in highly mixing regions by the methods explained in section \ref{sec:theory}. Setting $K=50$ mutual neighbors is enough to obtain sufficiently smooth plots and resolve boundary layers where tracers get sticky.

For the $\Ra$ considered, we expect the center of each convective cell to behave as a Lagrangian vortex or, by definition, as a nested family of elliptic LCSs surrounding a unique local maximum of the LAVD field, but we do not get that picture from our plots (see figure \ref{fig:fig3}). First, LAVD as a field provides a complementary picture from stretching measures, highlighting rotationally coherent structures. However, maximum values inside each of these regions appear near its boundary and not near its supposed center of rigid rotation. This happens for every $\eta$ except for the smallest (hence the thinnest annulus), where curvature effects are less noticeable and convective rolls are symmetric. Thus, in every other case, if we look for local maxima of the LAVD field that lie within its 60th percentile, only points near the boundary of each convective cell are detected. For these points, in most cases we are not able to find closed and convex LAVD contours that allow us to claim the detection of a significant Lagrangian vortex. 

As this definition is too restrictive for us to correctly determine the underlying dynamics, we relaxed the convexity condition, that is to say, we account for some \textit{convexity deficiency} defined as the relative difference in area between each closed curve and its convex hull. We set this value to 0.25 and we immediately start to obtain Coherent Lagrangian Swirls (CLSs), as first reported and defined by \cite{Vera2019}. In figure \ref{fig:swirl_advected}, we present an illustrative example of a detected CLS inside a convective cell. First, we initialize 3000 red tracers along the detected CLS and another 6000 tracers inside (purple) and outside (white) the boundary. Then, we let the colored tracers  evolve with the flow map for $T=31$ and plot their initial and final positions over the corresponding LAVD fields. In light of figure \ref{fig:swirl_advected}, we see that the CLS develops filaments as it shrinks toward the center of the cell, but these filaments do not undermine the global structure. As expected, after such a long time in our free-fall time units, it still behaves as a mixing barrier: not a single purple tracer has left the CLS at any time; these coherent swirls are minimally dispersive, as diagnosed also by the node degree. 

A relevant observation is the following: we are placing a given number of tracers on an initially material closed curve, thus we do not expect the resulting area enclosed by them to be preserved over time, that is, we are not computing a contour advection. Thus, the incompressibility of the flow (or the area-preserving property) is not at risk due to this kind of dynamics.  

In fact, this behavior is a reflection of the resistance to deformation of tracers within the coherent swirl. Their large relative bulk rotation values (or LAVD) strongly oppose the stretching or folding induced by the corresponding hot or cold plume. Enhanced by the radial asymmetry of the flow and, thus, the shear stress between the cell and the cold plume, tracers seem to spiral inward reminiscing an attracting fixed point inside the cell.

Moreover, from a physical perspective, we can plot an angular velocity profile passing through the detected CLS center (see figure \ref{fig:swirl_advected}). The boundaries of this counterclockwise CLS (vertical dashed lines in figure \ref{fig:swirl_advected}) divide two regions where angular velocity changes its tendency. Thus, we can think of a relatively similar behavior as the idealized Rankine vortex, the starting point of some wind distribution models in hurricanes see \cite{acheson1990elementary}. Consequently, this swirling dynamics becomes clear even by first principles: the CLS core behaves as a rotating rigid body surrounded by an irrotational outer flow (a Couette flow).

We now have a complete Lagrangian picture of the generic flow dynamics and will shift our focus on special dynamical phenomena.

\subsection{Coupling of heat flux and the FTLE/node degree}
\label{sec:heat_flux}

Applying the sliding window approach described in section \ref{subsec:sliding_window}, we expect to identify relevant flow dynamics by inspecting spatially averaged \textit{f}-FTLE or normalized node degree curves computed over  windows of length $T=1$ and plotted with respect to the beginning or center of each time interval (see \eqref{eq:FTLE_mean}, \eqref{eq:node_mean} and figure \ref{fig:sliding_window}). In this sense, we suggest that relevant fluid phenomena, such as the splitting or merging of plumes or the oscillatory motion of the base flow, should leave a trace on these trajectory-based methods. For comparison, we compute an instantaneous Nusselt curve, which encodes information about heat transfer evolution in the flow. For sufficiently long time, we expect that all these curves reach a stationary motion, meaning that the flow has stabilized around some equilibrium.

\begin{figure}
        \centering
        \includegraphics{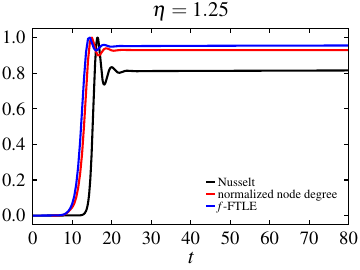}
        \includegraphics{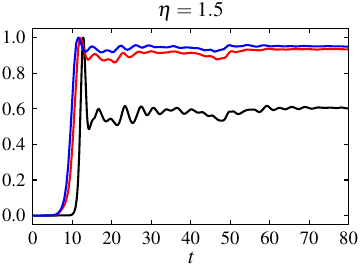}\\
        
        \includegraphics{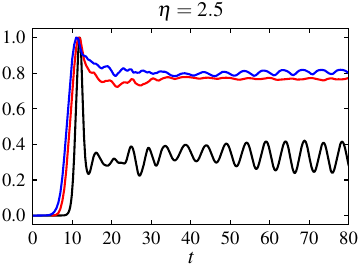}
        \includegraphics{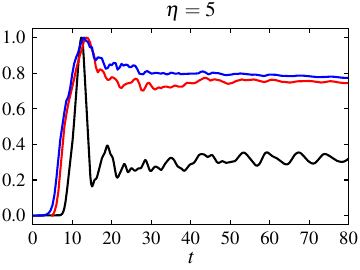}
    \caption{Nusselt, \textit{f}-FTLE and normalized node degree curves computed using the sliding window approach for all annular widths and $\Ra=1 \cdot 10^6$. All curves are normalized to take values from 0 to 1.}
    \label{fig:Nu_fftle}%
\end{figure}

In figure \ref{fig:Nu_fftle}, we present some selected cases for a fixed $\Ra$. The kinematic measures of mixing ---\textit{f}-FTLE and normalized node degree--- mostly replicate the radial heat flux dynamics represented by the Nusselt curve, especially when the steady state sets in, thus a single dynamical regime is driving the flow. We say that they are \textit{coupled}. Note that there is an offset between the different techniques that persists for all annular widths. On the one hand, the Nusselt curve encodes the instantaneous spatially-averaged radial heat flux, while the \textit{f}-FTLE and the node degree curves encode future (forward) and a combination of past and future (backward and forward) information about trajectories, respectively. Hence, the \textit{f}-FTLE curve is supposed to be ahead of the Nusselt curve and, on the contrary, the node degree will be lagged behind.

Therefore, we should use these time evolution curves of the kinematic measures of mixing as tools that describe, to a large extent, an instantaneous radial heat flux value; but we cannot use them to predict dynamical transitions (e.g., upcoming bifurcations), as we include future information that is strictly not yet available. 

On the other hand, in figure \ref{fig:Nu_fftle} we see that the system reaches different asymptotical behaviors: from steady motion to oscillatory modes. This spectrum of dynamical configurations shows that changes in the annular geometry affect the long-term flow configurations of the system. The remainder of this section is devoted to a more detailed study of the different regimes shown in these curves.

\subsubsection*{Relevant dynamical transitions}

\begin{figure}
        \centering
        \includegraphics[width=0.8\textwidth]{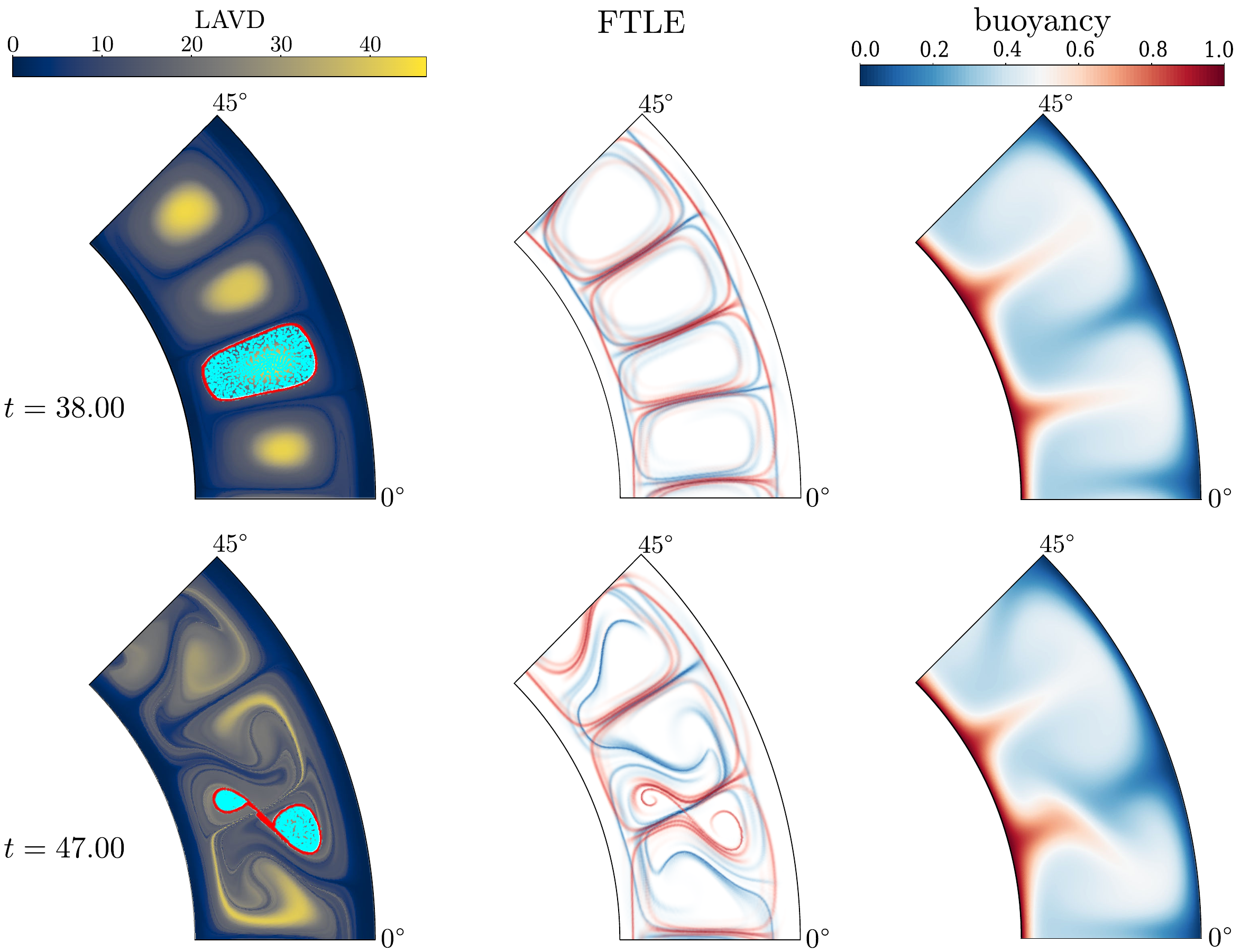}
        
    \caption{LAVD-based coherent set splitting event for $\eta=1.5$, $\Ra=1 \cdot 10^6$. Left column: LAVD-based coherent set detected with $t_0 = 30$, $T=8$ over its corresponding LAVD contour and a further advected moment with $t_0=47$, $T=8$. Tracers are initialized at $t=38.0$: the red ones over the CLS boundary and the cyan ones remain in the interior subset. Middle column: \textit{f}-FTLE (blue) and \textit{b}-FTLE (red) contours plotted at the same time stamps ($T=3$, adapted to the duration of the event). Both techniques are filtered to highlight values over its 70th percentile. Right column: buoyancy frames at the same time stamps.}
    \label{fig:fig6}%
\end{figure}

First of all, let us focus on the curves shown in figure \ref{fig:Nu_fftle} for $\eta = 1.5$, $\Ra = 1 \cdot 10^6$. When $t\in [40,50]$, we see a sudden drop of the radial heat flux that leaves a trace on the kinematic structure of mixing. 
By looking at instantaneous buoyancy simulation frames (see the right column in figure \ref{fig:fig6}), we recognize an unstable region where some plumes are interacting strongly with each other, even though in the rest of the domain the situation seems to be more stable. To deeply understand the underlying dynamics in this region, Lagrangian methods can be very useful. We perform a LAVD computation before this instability sets in ($t_0=30$, $T=8$) and we are able to find an interesting CLS under the same assumptions as in section \ref{sec:flow_field} (see red tracers in figure \ref{fig:fig6}). At $t=38$, we initialize tracers along the previous CLS, both over its boundary (red) and on its interior (cyan). By letting these tracers evolve with the flow, we find that around $t=47$ it splits into two sets that quickly spread out, losing coherence. In the hypothetical situation of an autonomous (i.e. time-independent) velocity field, we may heuristically relate this to an initially elliptic structure that has been perturbed and, all of a sudden, some hyperbolic saddle point has appeared exactly at the center of a previous convective cell, where high bulk rotation values were acting as a mixing barrier. Physically, the rigid body rotation inside the cell has become unstable, and as a result, two hot ascending plumes have merged. This interpretation is supported by the computation of both \textit{f}-FTLE and \textit{b}-FTLE overlapped on the same plot with $T= 3$ in order to focus on the event (see figure \ref{fig:fig6}). The resulting repelling and attracting LCSs evolve with time until they are rearranged into a configuration that will remain stable for the rest of the simulation. 

\begin{figure*}
        \centering
        \raisebox{0.2cm}{\includegraphics[width=0.5\textwidth]{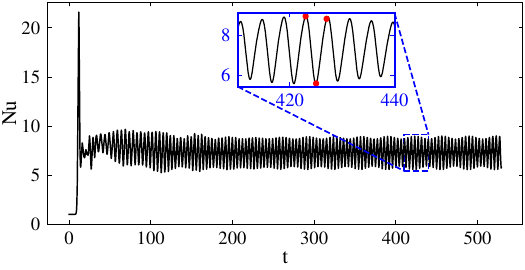}}
        \includegraphics[width=0.4\textwidth]{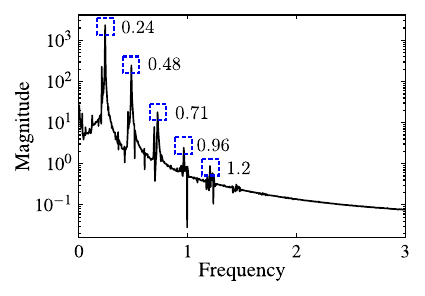}
        \includegraphics[width=0.8\textwidth]{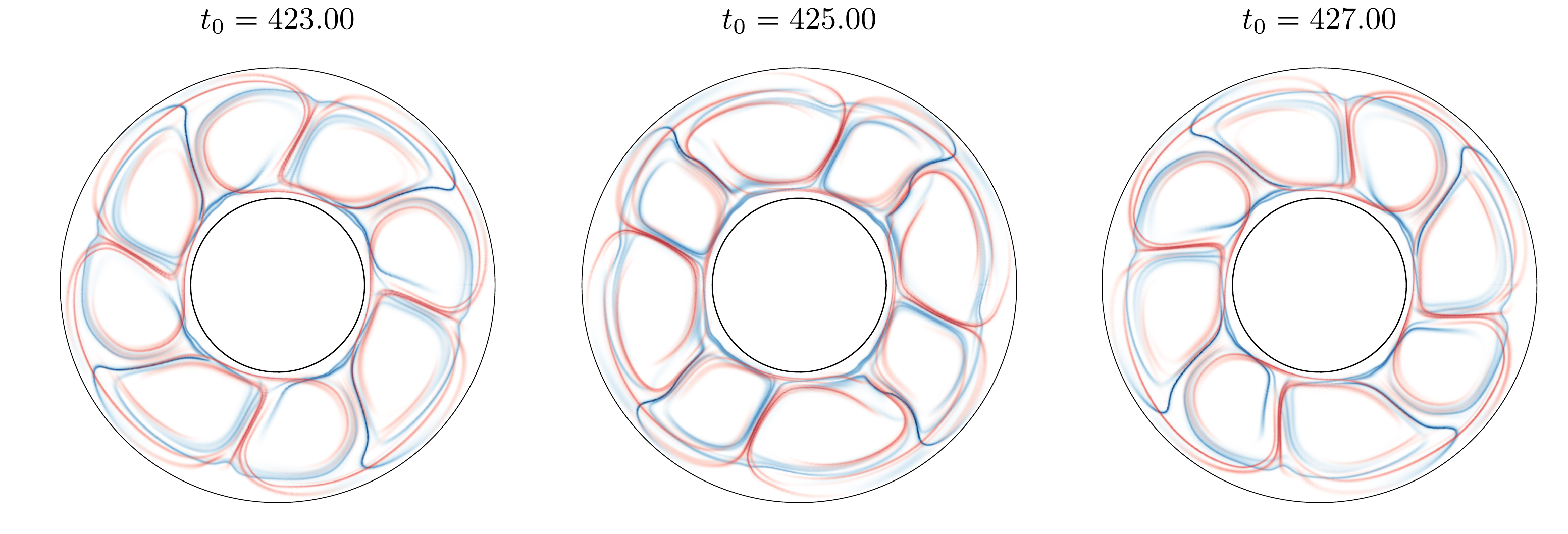}
        
    \caption{Long-term oscillating dynamics in the case $\eta=2.5$, $\Ra=1 \cdot 10^6$. Above: Nusselt number temporal evolution, the steady state behaves as a wave packet (left). The Fourier transform of this steady state shows different harmonics of a main peak (right). Below: strong LCS candidates for the \textit{f}-FTLE (blue) and \textit{b}-FTLE (red) fields with $T= 3$ plotted at the same red dots as the left panel above. FTLE values are filtered as in figure \ref{fig:fig6}.}
    \label{fig:Nu25_1e6}%
\end{figure*}

From all the different configurations simulated, there is another relevant kind of long-term dynamics that leaves a trace both on heat flux and on Lagrangian techniques and, therefore, can be studied in detail with our methods. It is the oscillatory motion of the plumes, which is most visible in the case $\eta = 2.5$, $\Ra= 10^6$ (see figure \ref{fig:Nu25_1e6}. Here, we see that the asymptotical behavior of the radial heat flux long after the turnover time is a modulated wave packet. We perform a Fourier transform of this Nusselt number signal from $t=200$ onward and we get a main frequency and its first four harmonics, meaning that these oscillations are a result of a weakly nonlinear instability of the base flow, i.e., all plumes oscillate at the same rate. We perform the same FTLE computation as for the previous splitting event, in order to check if our attracting and repulsing LCSs replicate these oscillations. The result is shown in figure \ref{fig:Nu25_1e6} and in fact, it can also be noticed in figure \ref{fig:fig3}. Based on its morphology, there appear to be two different kinds of convective cells ---we have one thinner cell in the middle of each two wider ones--- which are clearly exchanged every two consecutive maxima. Either way, by the Fourier analysis we can assure that every cell oscillates in phase.

A related phenomenon is discussed by \cite{BHADRA2025126703} as zonal flow. We report periodic oscillations of the convective rolls in some simulations, but we have not seen uniform motion of the plumes in one direction. Similar drifting or zonal-flow behavior has also been reported in rotating thermal convection (e.g., \cite{arantxa1998,arantxa2003}). In fact, in our simulations almost every oscillatory motion tends to vanish over sufficiently long integration times, with the exception of the case presented above. To the best of our knowledge, we have not found other reports of this type of behavior in two-dimensional thermal convection without imposed rotation.

\subsection{Approaching boundary layer detection from Lagrangian techniques}
\label{sec:TBLs}
\begin{figure}
        \centering
        \includegraphics[width=0.48\textwidth]{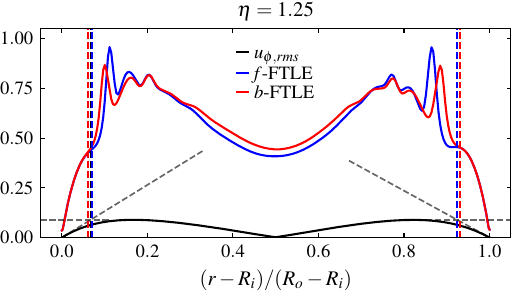}
        \hfill
        \includegraphics[width=0.48\textwidth]{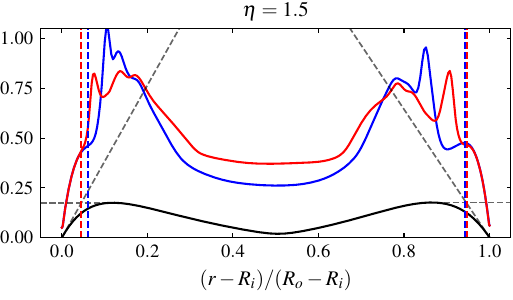}
        \vspace{0.1cm}
        \includegraphics[width=0.48\textwidth]{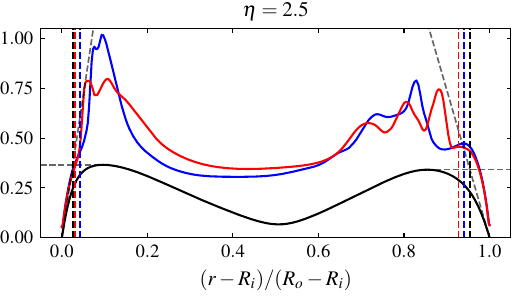}
        \hfill
        \includegraphics[width=0.48\textwidth]{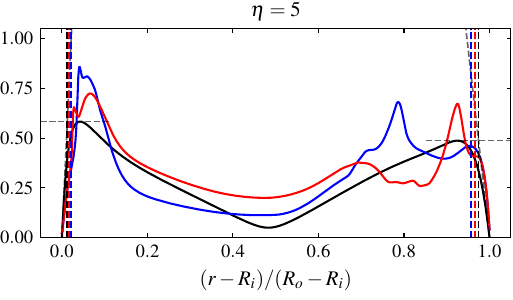}
    \caption{Examples of the proposed Lagrangian UBL estimation method using FTLE radial profiles computed with $T= 3$ for all $\eta$ and $\Ra=1\cdot 10^6$. Vertical dashed lines show the limits of the corresponding UBL width estimate: \textit{f}-FTLE (blue), \textit{b}-FTLE (red) and slope (black). The other dashed lines are the auxiliary ones needed for the slope method.}
    \label{fig:UBLs}%
\end{figure}

An important issue to consider when trying to understand turbulent heat and momentum transport in convective problems is providing a good characterization of the boundary layer regions see \cite{Tilgner_1996, Breuer2004, Shiskina2010, Gastine_Wicht_Aurnou_2015, Bhadra_Shishkina_Zhu_2024, BHADRA2025126703}. As \textit{no-slip} boundary conditions are imposed, a monotonic behavior of fluid properties ---velocity and buoyancy--- is expected near the walls. This splits the fluid motion into a bulk flow surrounded by a thin layer of fluid characterized by almost motionless Lagrangian tracers. This observation, along with the implementation of different Lagrangian methods in the previous sections, motivates us to propose a new boundary layer estimation method based on finite-time tracer trajectories and, in particular, on FTLE radial profiles. By looking at figures \ref{fig:fig6} and \ref{fig:Nu25_1e6}, we might expect that after the last hyperbolic LCS candidate (or mixing barrier) detected by the FTLE near each boundary, angular velocity starts to decrease monotonically, marking the limit of the so-called \textit{kinetic} or \textit{viscous} boundary layer region (usually denoted as UBL). Of course, in this geometry we have an inner UBL and an outer UBL, whose widths are $\lambda_U^i$ and $\lambda_U^o$, respectively. 

\begin{figure}
        \centering
        \includegraphics{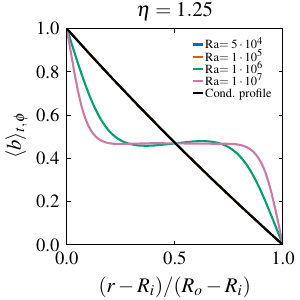}
        \includegraphics{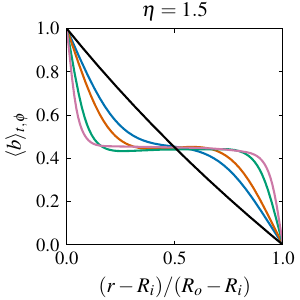}\\
        
        \includegraphics{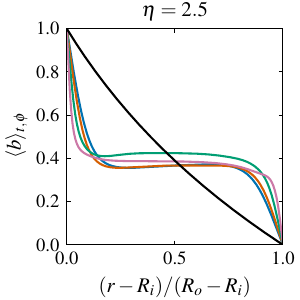}
        \includegraphics{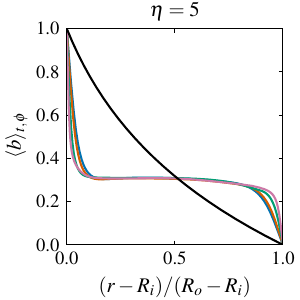}
    \caption{Time and angularly-averaged buoyancy radial profiles.}
    \label{fig:buoy_profiles}%
\end{figure}

In the literature, we find several ways of estimating UBL widths, all based on different assumptions on angular velocity decay near the walls or, in our broader context, on Eulerian methods. In \cite{Gastine_Wicht_Aurnou_2015}, two different definitions are compared, showing distinct UBL width values. In particular, they assess both methods in relation to the classical Grossmann-Lohse (GL) theory see \cite{GROSSMANN_LOHSE_2000} which, for unity Prandtl and relatively high Reynolds number, states that UBL width must be approximately equal to its thermal counterpart, the thermal boundary layer (TBL). Under these assumptions, they accept that the most precise UBL (and TBL) estimation method is the slope method, which is widely used in thermal convection studies such as \cite{Breuer2004, AHLERS_2006, Shiskina2010, Bhadra_Shishkina_Zhu_2024, BHADRA2025126703}. 

In this framework, $\lambda_{U, \text{slope}}^{i/o}$ are defined as the radial distances from the closest wall to the intersection of a linear fit of the angular velocity root mean square ($u_{\phi, rms}$) radial profile at the walls and a horizontal line through the closest maximum. In figure \ref{fig:UBLs}, we show some examples of these $u_{\phi, rms}$ radial profiles, where the intersection of the corresponding dashed lines show the exact point that leads to this UBL width definition. Similarly, TBL widths $\lambda_{T, \text{slope}}^{i/o}$ are defined as the radial distances from the closest wall to the intersection of a linear fit of the temperature radial profile at the walls and the horizontal line through the temperature at mid-radius. Figure \ref{fig:buoy_profiles} shows temperature radial profiles of every simulation performed along with the conductive profile imposed. We report strong asymmetry in the mean temperature values at radial mid-point at larger $\eta$ and pronounced temperature gradients at mid-points also for larger $\eta$. This agrees with \cite{BHADRA2025126703} simulations. However, once again we see that the singular case $\eta=2.5$ and $\Ra=1\cdot 10^6$ stands out for disturbing the trend: even though no noticeable effect is seen for changes in $\Ra$, this zonal flow increases mean temperature at mid-point. And this is linked to a weaker temperature drop in the inner boundary, followed by a slight temperature gradient before reaching the mid-point.

\begin{figure}
    \centering
    \begin{minipage}[c]{0.43\textwidth}
        \includegraphics[width=0.75\textwidth]{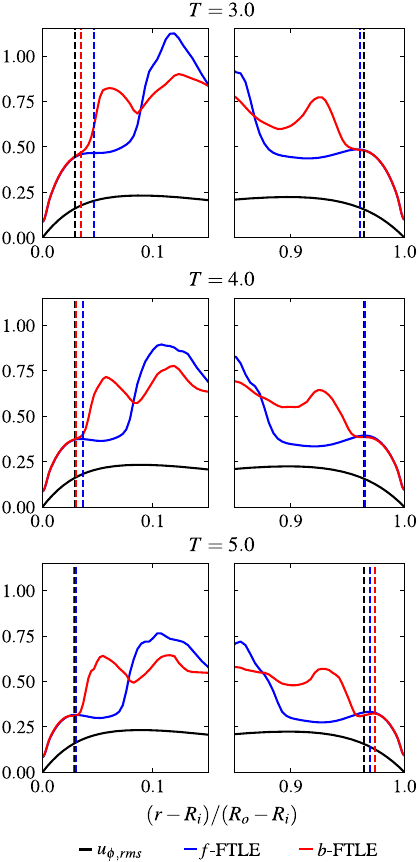}
    \end{minipage}
    \hfill
    \begin{minipage}[c]{0.55\textwidth}
        \includegraphics[width=0.95\textwidth]{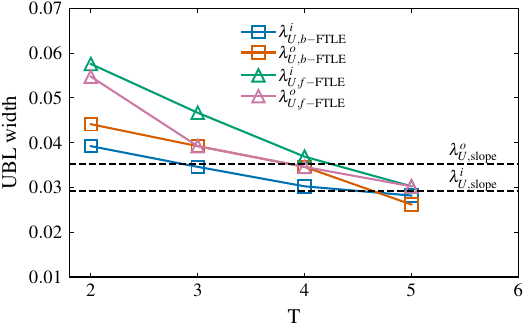}\\[0.5cm]
        \includegraphics[width=0.95\textwidth]{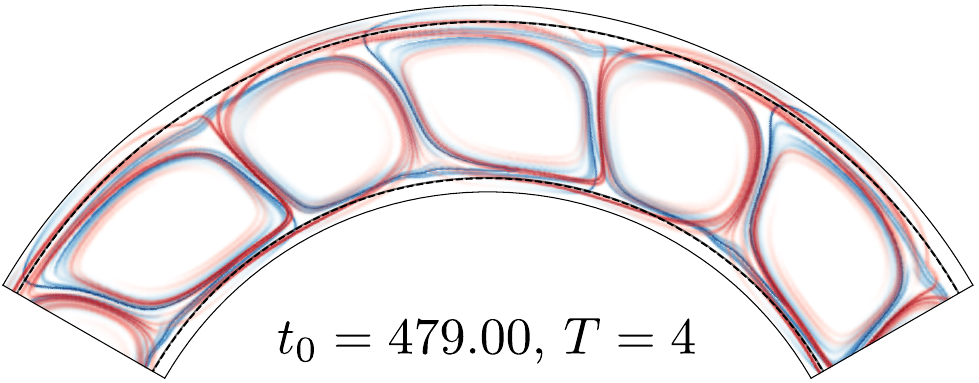}
    \end{minipage}
    
    \caption{UBL estimates for different integration times in the case $\Ra = 1 \cdot 10^7$, $\eta=1.5$. Dashed lines show the estimates obtained with the three different methods, with colors matching their corresponding solid lines. Lower right panel shows UBL estimates using the slope method (dashed) and the \textit{b}-FTLE method (solid, grey) over both filtered FTLE fields. The difference is almost imperceptible.}
    \label{fig:UBLS_T}
\end{figure}

Back to our Lagrangian description, we compute angular averages of the FTLE at the end of each simulation, that is, when the steady state is reached. Using these FTLE radial profiles, we expect to capture the exact radius value from which tracers are ejected from the viscous boundary layer for a given finite-time interval, meaning that they jump into the bulk flow in such a time. From this radial point, a \textit{b}-FTLE-based hyperbolic LCS (presumably attracting) is always followed closely by a \textit{f}-FTLE-based one (presumably repelling). This is the situation shown in figure \ref{fig:UBLs}, where highly symmetric profiles for lower $\eta$ tend to be displaced to inner radii when $\eta$ is increased. This is consistent with the increasing radial asymmetry argument we have used before. We also see that FTLE profiles are monotonous near the boundaries until a change in curvature is reached before/after the innermost/outermost LCS. This appears to be either a maximum (critical point) or an inflection (critical point of the first derivative). This is exactly the point we have been looking for: the last radius for which a Lagrangian tracer is subject to the influence of the UBL (for a fixed $T$). Thus, we shall define our UBL inner/outer finite-time limit estimate as the innermost/outermost FTLE radial profile maximum/inflection from which tracers are dragged toward the bulk flow. If a stationary regime is assumed, this finite-time estimate should be robust under time shifts. Therefore, we propose two different ways of estimating UBL width from finite-time Lagrangian methods: one using backward trajectories via \textit{b}-FTLE profiles ($\lambda_{U, \textit{b}\text{-FTLE}}^{i/o}$) and another one using forward information via \textit{f}-FTLE profiles ($\lambda_{U, \textit{f}\text{-FTLE}}^{i/o}$). Some examples are shown in figure \ref{fig:UBLs} for a fixed $\Ra$.

\subsubsection*{Fixing an appropriate integration time for UBL detection}

As noted above, our Lagrangian finite-time UBL estimate is sensitive to the length of the time interval ---as this defines the whole dynamics--- but not to the chosen $t_0$ (as long as the stationary state is reached). Thus, we need to carefully unveil the appropriate integration time $T$ that correctly captures the underlying dynamics of this convective flow. We will use the slope method estimate for comparison.

Let us consider the case $\eta=1.5$, $\Ra=1\cdot 10^7$ and compute FTLE radial profiles with $T$ ranging from 2 to 5 (see figure \ref{fig:UBLS_T}). As we assume sufficient regularity of the flow map \eqref{eq:deformaciones}, we systematically compute $\lambda^{i/o}_{U, \textit{f}\text{-FTLE}}$, $\lambda^{i/o}_{U, \textit{b}\text{-FTLE}}$ simply by finding the innermost/outermost radial value that cancels the first derivative of the FTLE radial profile (which always turns out to be a maximum) or, correspondingly, the second derivative (an inflection). Results shown in figure \ref{fig:UBLS_T} tell us that FTLE estimates monotonically decrease when $T$ is increased, i.e., when the FTLE fields become saturated. Optimal UBL estimates are obtained when $T \in [3, 5]$. We can make two critical observations. First, \textit{b}-FTLE estimates maintain their outer/inner gap until something happens near the outer boundary between $T=4$ and $T=5$, where the inner UBL gets thinner than the outer one. This would contradict the physical model proposed by \cite{Bhadra_Shishkina_Zhu_2024}. Looking closer at the profiles we see that the inflection detected with $T=4$ becomes a maximum with $T=5$, explaining the sudden gap change. Thus we will not assume that any $T \geq 5$ may properly describe UBL regions. On the other hand, \textit{f}-FTLE estimates also provide an inverse gap for lower $T$, which can be explained as a detection of an inflection with $T=3$ that becomes a maximum with $T=5$. This happens with forward information in some other cases (see table \ref{table:datas}) so, in general, we stick to backward information as the one that gives physically reliable results.

\begin{figure*}
    \centering    \includegraphics[width=0.49\textwidth]{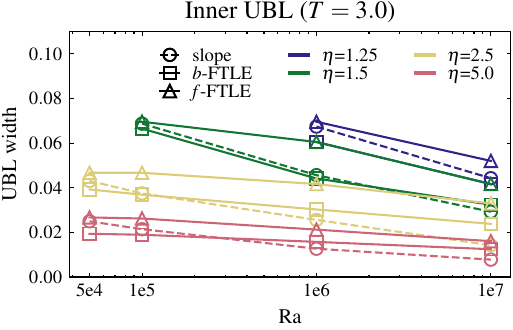}
    \hfill
    \includegraphics[width=0.49\textwidth]{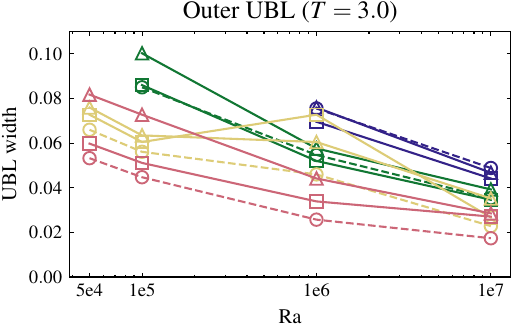}
    \includegraphics[width=0.49\textwidth]{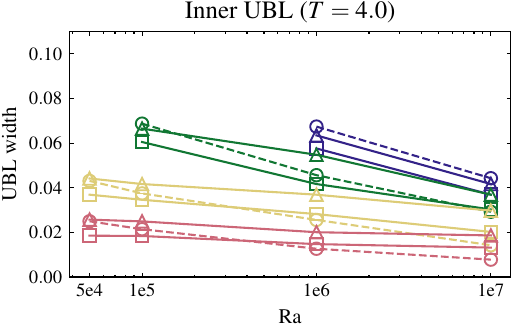}
    \hfill
    \includegraphics[width=0.49\textwidth]{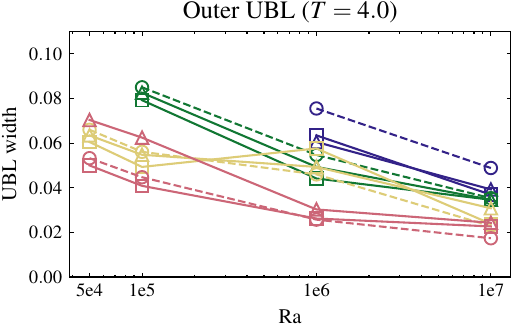}
    \caption{Summary of the UBL estimates computed with $T=3$ (above) and $T=4$ (below) using the methods explained in the text. We have separated inner and outer BLs for clarity. Values shown in the case $\eta=5$, $\Ra=1 \cdot 10^7$ are inconclusive (see caption in table \ref{table:datas} for details).}
    \label{fig:UBL_summary}%
\end{figure*}

In figure \ref{fig:UBL_summary}, we show UBL widths obtained using the three approaches described for the two optimal integration times proposed, $T=3$ and $T=4$. With increasing $\Ra$, UBLs get thinner. This trend is generally verified by every technique, except for the outer UBL in the particular case $\eta=2.5$, $\Ra=1\cdot 10^6$. The three methods see a noticeable disagreement there, overestimating the general trend. This was the case discussed in section \ref{sec:heat_flux}, where a zonal flow was detected. This instability of the plumes near the outer boundary leads to unexpected UBL widths. Lagrangian methods are successful in describing this phenomenon. On the other hand, $T=3$ provides closer results for lower $\eta$, but $T=4$ seems to be better for larger $\eta$. This suggests that any change in the geometry affects the finite-time dynamics. Consequently, the integration time must be adjusted for optimal results.

Either way, we conclude that \textit{b}-FTLE (hence locally most attracting structures) gives a fairly good estimate of the viscous boundary layers limits, as long as we are able to properly identify the time scale of the underlying dynamics, which depends sensibly on the parameters involved. Results are summarized in table \ref{table:datas} for $T=3$ and $T=4$.

\section{Conclusions and outlook}
\label{sec:discussion}
Long-lived coherent structures in the Lagrangian frame of reference provide crucial insights on fluid mixing and transport with strong applications in many contexts. In the present work, we investigate how the large scale circulation of the classical Rayleigh-Bénard convection problem is affected by an annular geometry under the influence of radial gravity. By heating the inner plate of this simple 2D annulus, we are able to assess the asymmetry due to the non-zero intrinsic curvature. As the aspect ratio $\eta$ of the annulus increases, the initially symmetric convective rolls become narrower near the inner plate and wider near the outer plate. Consequently, the radial velocity of the hot ascending plumes exceeds that of the cold descending ones. This behavior leads to a shear stress between rolls and plumes whose net effect is to make Lagrangian trajectories, i.e. passive tracers that follow the velocity field, spiral inwards and introduce some oscillatory modes that we relate to weakly nonlinear instabilities. The Rayleigh numbers covered in this study do not allow for fully developed turbulent bulk flows, thus such structures remain stable for long times. A deep understanding of the flow kinematics in the context of basic hydrodynamical instabilities is essential for the ultimate goal of developing predictive models for the behavior of complex real-world flows in geophysical and astrophysical settings. 

From this starting point, we were able to characterize convective plumes as hyperbolic highways of mixing in the flow. Maximum curves of the \textit{f}-FTLE field were heuristically related to hot ascending plumes (potentially repulsive) and maximum curves of the \textit{b}-FTLE to cold descending plumes (potentially attractive). On the other hand, convective rolls were associated to LAVD-based Coherent Lagrangian Swirls (CLSs), mainly due to the swirling motion described above and, hence, strongly non-mixing subsets. These characterizations were supported by node degree computations. Moreover, we computed a sliding window scheme of the Lagrangian stretching tools in order to successfully capture radial heat transport, as measured by the Nusselt number. This novel instantaneous spatially-averaged FTLE and normalized node degree value allow us to infer the appearance of distinct dynamical transitions associated to intermittency by just using sets of trajectories. The resemblance seems to work better for small values of $\eta$, i.e., thinner annuli. This is due to the distinctive flow field produced by curvature effects. From this, we characterized a splitting event and a singular librating mode of the plumes, both leaving traces on radial heat flux and on Lagrangian metrics. 

Motivated by the boundary layer asymmetry discussed and also physically modeled by \cite{Bhadra_Shishkina_Zhu_2024, BHADRA2025126703}, we used FTLE radial profiles to propose a new viscous boundary layer (UBL) width estimation criteria based on the innermost/outermost critical point of this azimuthally-averaged Lagrangian stretching measure. Results were compared to the widely used Eulerian slope method that uses the decay slope of the $u_{\phi, rms}$ radial profile. From the Lagrangian perspective, we obtained a good agreement with the slope method, subject to a strict analysis of the finite time interval of the underlying dynamics. In our case, $T=3$ works better for lower $\eta$ and $T=4$ for larger $\eta$. If $T$ is further increased, UBLs get underestimated, providing nonphysical results. Furthermore, \textit{b}-FTLE gives better estimates of the UBLs widths than the \textit{f}-FTLE, probably due to the (not-necessarily) attracting nature of backward trajectories.

In the future, this project can be extended in two plausible ways. On the one hand, we plan to extend our setup to a 3D cylindrical shell as we would like to have some more insight on the three-dimensional structure of the flow field. This would also allow us to incorporate rotation. If thin-disk dynamics are considered, this would move the study towards the astrophysical application mentioned in the introduction. On the other hand, applying other Lagrangian techniques and computational methods for the detection of coherent sets could be explored; one option is to perform the evolutionary clustering scheme developed by \cite{KPG_RBC2022}. In this regard, we also expect to reach larger $\Ra$, as characterizing slightly turbulent bulk flows and BLs would be of interest for our approach. Once we have our flow dynamics classified by coherent motions, we will be able to assess how they impact Lagrangian transport and mixing by themselves.

\textbf{Acknowledgements} {L. Álamo and J. Curbelo thank Gabriel Meletti (Universitat Politècnica de Catalunya) for helpful discussions on the initial results. K. Padberg-Gehle acknowledges hospitality and support by Centre de Recerca Matem\`atica and Departament de Matem\`atiques, Universitat Polit\`ecnica de Catalunya (UPC) Barcelona, Spain.}

\textbf{Funding} {L. Álamo's research is funded by the Departament de Recerca i Universitats through a EU Joan Oró 2025 FI-100704 grant. L. Álamo and J. Curbelo acknowledge support from  the Agencia Estatal de Investigación through the grant  CNS2023-144360 funded by the “European Union NextGenerationEU/PRTR” and PID2024-159949NA-I00. She also thanks the Severo Ochoa and María de Maeztu Program for Centers and Units of Excellence in R\&D (CEX2020-001084-M) and the Fundación Ramón Areces. K. Padberg-Gehle's research has been partially supported by the Deutsche Forschungsgemeinschaft (DFG, German Research Foundation) -- SFB 1615 -- 503850735, and by the Centre de Recerca Matem\`atica, Barcelona, Spain.}

\textbf{Declaration of interests} {The authors report no conflict of interest.}

\textbf{Author ORCIDs} {L. Álamo https://orcid.org/0009-0009-6500-6896; J. Curbelo, https://orcid.org/0000-0002-7276-0127; K. Padberg-Gehle, https://orcid.org/0000-0002-1761-213X}

\appendix

\section{Summary of simulation results}\label{appA}
\noindent In table \ref{table:datas}, we show details of the simulations performed with \texttt{Dedalus3} and the UBL results plotted in figures \ref{fig:UBLs}, \ref{fig:buoy_profiles}, \ref{fig:UBLS_T} and \ref{fig:UBL_summary}, which complements the text in section \ref{sec:TBLs}.
\begin{table}
\centering
\resizebox{\textwidth}{!}{\begin{tabular}{@{}cccccccc@{}}
\midrule
    $\Ra$ & $\Nu$ & $n_r \times n_\phi$ & $T$ & $\lambda^i_{U, \textit{b}\text{-FTLE}}/\lambda^o_{U, \textit{b}\text{-FTLE}}$ &  $\lambda^i_{U, \textit{f}\text{-FTLE}}/\lambda^o_{U, f\text{-FTLE}}$ & $\lambda^i_{U, \text{slope}}/\lambda^o_{U, \text{slope}}$ & $\lambda^i_{T, \text{slope}}/\lambda^o_{T, \text{slope}}$ \\ \midrule
\multicolumn{8}{c}{$\boldsymbol{\eta = 1.25}$} \\
$5 \cdot 10^4$ & - & $256 \times 3072$ &  & - & - & - & - \\
$1 \cdot 10^5$ & - & $256 \times 3072$ &  & - & - & - & - \\
\multirow{2}{*}{$1 \cdot 10^6$} & \multirow{2}{*}{2.81} & \multirow{2}{*}{$256 \times 3072$} & 3 & 0.060/0.070 & 0.070/0.076 & \multirow{2}{*}{0.067/0.076} & \multirow{2}{*}{0.168/0.187} \\
               &   &                      & 4 & 0.058/0.063 & \textit{0.063/0.060} &  &             \\
\multirow{2}{*}{$1 \cdot 10^7$} & \multirow{2}{*}{5.29} & \multirow{2}{*}{$256 \times 3072$} & 3 & 0.042/0.044 & \textit{0.052/0.047} & \multirow{2}{*}{0.044/0.049} & \multirow{2}{*}{0.090/0.099} \\
               &   &                      & 4 & 0.037/0.037 & \textit{0.042/0.039} &  &             \\
\multicolumn{8}{c}{$\boldsymbol{\eta = 1.5}$} \\
$5 \cdot 10^4$ & - & $256 \times 3072$ &  & - & - & - & - \\
\multirow{2}{*}{$1 \cdot 10^5$} & \multirow{2}{*}{2.43} & \multirow{2}{*}{$256 \times 3072$} & 3 & 0.066/0.086 & 0.070/0.100 & \multirow{2}{*}{0.069/0.085} & \multirow{2}{*}{0.184/0.225} \\
               &   &                      & 4 & 0.060/0.079 & 0.066/0.083 &  &             \\
\multirow{2}{*}{$1 \cdot 10^6$} & \multirow{2}{*}{4.72} & \multirow{2}{*}{$256 \times 3072$} & 3 & 0.044/0.052 & \textit{0.060/0.058} & \multirow{2}{*}{0.046/0.055} & \multirow{2}{*}{0.096/0.114}\\
               &   &                      & 4 & 0.042/0.044 & \textit{0.055/0.049} &  &             \\
\multirow{2}{*}{$1 \cdot 10^7$} & \multirow{2}{*}{7.94} & \multirow{2}{*}{$256 \times 3072$} & 3 & 0.032/0.035 & \textit{0.042/0.039} & \multirow{2}{*}{0.029/0.035} & \multirow{2}{*}{0.056/0.069} \\
               &   &                      & 4 & 0.030/0.035 & \textit{0.037/0.035} &  &             \\
\multicolumn{8}{c}{$\boldsymbol{\eta = 2.5}$} \\
\multirow{2}{*}{$5 \cdot 10^4$} & \multirow{2}{*}{3.92} & \multirow{2}{*}{$256 \times 1800$} & 3 & 0.039/0.073 & 0.047/0.076 & \multirow{2}{*}{0.043/0.066} & \multirow{2}{*}{0.099/0.143} \\
               &   &                      & 4 & 0.037/0.060 & 0.044/0.063 &  &             \\
\multirow{2}{*}{$1 \cdot 10^5$} & \multirow{2}{*}{4.76} & \multirow{2}{*}{$256 \times 1800$} & 3 & 0.037/0.060 & 0.047/0.063 & \multirow{2}{*}{0.037/0.056} & \multirow{2}{*}{0.081/0.118}  \\
               &   &                      & 4 & 0.035/0.049 & 0.042/0.055 &  &             \\
\multirow{2}{*}{$1 \cdot 10^6$} & \multirow{2}{*}{7.36} & \multirow{2}{*}{$256 \times 1800$} & 3 & 0.030/0.073 & 0.042/0.060 & \multirow{2}{*}{0.026/0.046} & \multirow{2}{*}{0.048/0.088} \\
               &   &                      & 4 & 0.028/0.058 & 0.037/0.049 &  &             \\
\multirow{2}{*}{$1 \cdot 10^7$} & \multirow{2}{*}{13.08} & \multirow{2}{*}{$512 \times 1800$} & 3 & 0.024/0.027 & 0.033/0.035 & \multirow{2}{*}{0.014/0.023} & \multirow{2}{*}{0.029/0.045} \\
               &   &                      & 4 & 0.020/0.024 & 0.030/0.031 &  &             \\
\multicolumn{8}{c}{$\boldsymbol{\eta = 5}$} \\
\multirow{2}{*}{$5 \cdot 10^4$} & \multirow{2}{*}{5.40} & \multirow{2}{*}{$256 \times 1800$} & 3 & 0.019/0.060 & 0.027/0.082 & \multirow{2}{*}{0.025/0.053} & \multirow{2}{*}{0.052/0.115} \\
               &   &                      & 4 & 0.019/0.050 & 0.026/0.070 &  &             \\
\multirow{2}{*}{$1 \cdot 10^5$} & \multirow{2}{*}{6.40} & \multirow{2}{*}{$256 \times 1800$} & 3 & 0.019/0.051 & 0.026/0.073 & \multirow{2}{*}{0.021/0.045} & \multirow{2}{*}{0.044/0.096} \\
               &   &                      & 4 & 0.018/0.041 & 0.025/0.062 &  &             \\
\multirow{2}{*}{$1 \cdot 10^6$} & \multirow{2}{*}{11.03} & \multirow{2}{*}{$768 \times 1800$} & 3 & 0.016/0.034 & 0.021/0.044 & \multirow{2}{*}{0.013/0.026} & \multirow{2}{*}{0.025/0.056} \\
               &   &                      & 4 & 0.015/0.026 & 0.020/0.030 &  &             \\
\multirow{2}{*}{$1 \cdot 10^7$} & \multirow{2}{*}{\textit{18.21}} & \multirow{2}{*}{$1440\times 1800$} & 3 & \textit{0.012/0.027} & \textit{0.016/0.028} & \multirow{2}{*}{\textit{0.008/0.017}} & \multirow{2}{*}{\textit{0.016/0.034}} \\
               &   &                      & 4 & \textit{0.013/0.023} & \textit{0.019/0.024} &  &             \\ \bottomrule
\end{tabular}
}
\caption{Summary table with simulation details and UBL results of section \ref{sec:TBLs}. The Nusselt number ($\Nu$) is a combined average in space and time over the steady motion. Boundary layer results are shown when they can be clearly identified with $T=3$ and $T=4$. For the sake of completeness, we also show TBL estimates obtained using the slope method. Values in italics show physical inconsistencies. In the case $\eta=5$ and $\Ra=1\cdot 10^7$, the steady motion is only partially reached due to high computational cost. Some \textit{f}-FTLE estimates show larger $\lambda^i_{U, \textit{f}\text{-FTLE}}$ than $\lambda^o_{U, f\text{-FTLE}}$, which is inconsistent, as explained in the text.}
\label{table:datas}
\end{table}

\printbibliography[heading=bibintoc]

@article{Dedalus_article,
  title = {Dedalus: A flexible framework for numerical simulations with spectral methods},
  author = {Burns, K. J. and Vasil, G. M. and Oishi, J. S. and Lecoanet, D. and Brown, B. P.},
  journal = {Phys. Rev. Res.},
  volume = {2},
  issue = {2},
  pages = {23-68},
  numpages = {39},
  year = {2020},
  month = {4},
  publisher = {American Physical Society},
  doi = {10.1103/PhysRevResearch.2.023068}
}

@article{Jarvis2025,
    doi = {10.21105/joss.07948},
    url = {https://doi.org/10.21105/joss.07948},
    year = {2025},
    publisher = {The Open Journal},
    volume = {10},
    number = {113},
    pages = {7948},
    author = {Jarvis, A. and Ross, S. D.},
    title = {NumbaCS: A fast {P}ython package for coherent structure analysis},
    journal = {Journal of Open Source Software} }

@book{Haller_2023,
    place={Cambridge}, 
    title={Transport Barriers and Coherent Structures in Flow Data: Advective, Diffusive, Stochastic and Active Methods},
    publisher={Cambridge University Press},
    author={Haller, G.},
    year={2023}
}

@article{Haller_2016_DPD,
    title = {Dynamic rotation and stretch tensors from a dynamic polar decomposition},
    journal = {Journal of the Mechanics and Physics of Solids},
    volume = {86},
    pages = {70-93},
    year = {2015},
    issn = {0022-5096},
    doi = {https://doi.org/10.1016/j.jmps.2015.10.002},
    author = {Haller, G.}
}

@article{Haller_2016_LAVD, 
    title={Defining coherent vortices objectively from the vorticity},
    volume={795},
    DOI={10.1017/jfm.2016.151},
    journal={Journal of Fluid Mechanics},
    author={Haller, G. and Hadjighasem, A. and Farazmand, M. and Huhn, F.},
    year={2016},
    pages={136–173}
}

@article{Ali_Weidman_1990,
    title={On the stability of circular {C}ouette flow with radial heating},
    volume={220},
    DOI={10.1017/S0022112090003184},
    journal={Journal of Fluid Mechanics},
    author={Ali, M. and Weidman, P. D.},
    year={1990},
    pages={53–84}
}

@article{Tagg_Weidman_1990,
    title={Linear stability of radially-heated circular {C}ouette flow with simulated radial gravity},
    volume={58},
    issue={3},
    DOI={10.1007/s00033-006-0066-5},
    journal={Zeitschrift für angewandte Mathematik und Physik},
    author={Tagg, R. and Weidman, P. D.},
    year={2007},
    pages={431-456}
}

@article{Bhadra_Shishkina_Zhu_2024,
    title={On the boundary-layer asymmetry in two-dimensional annular {R}ayleigh–{B}énard convection subject to radial gravity},
    volume={999},
    DOI={10.1017/jfm.2024.995},
    journal={Journal of Fluid Mechanics},
    author={Bhadra, A. and Shishkina, O. and Zhu, X.},
    year={2024},
    pages={R1}
}

@article{BHADRA2025126703,
    title = {Heat and momentum transfer in {R}ayleigh–{B}énard convection within a two-dimensional annulus under radial gravity},
    journal = {International Journal of Heat and Mass Transfer},
    volume = {241},
    pages = {126703},
    year = {2025},
    issn = {0017-9310},
    doi = {https://doi.org/10.1016/j.ijheatmasstransfer.2025.126703},
    author = {A. Bhadra and O. Shishkina and X. Zhu}
}

@article{KPG2019,
    author = {Banisch, R. and Koltai, P. and Padberg-Gehle, K.},
    title = {Network measures of mixing},
    journal = {Chaos: An Interdisciplinary Journal of Nonlinear Science},
    volume = {29},
    number = {6},
    pages = {063125},
    year = {2019},
    month = {06},
    issn = {1054-1500},
    doi = {10.1063/1.5087632},

}

@article{KPG2017,
    AUTHOR = {Padberg-Gehle, K. and Schneide, C.},
    TITLE = {Network-based study of {L}agrangian transport and mixing},
    JOURNAL = {Nonlinear Processes in Geophysics},
    VOLUME = {24},
    YEAR = {2017},
    NUMBER = {4},
    PAGES = {661--671},
    DOI = {10.5194/npg-24-661-2017}
}

@article{KPG_RBC2021,
  title = {Lagrangian heat transport in turbulent three-dimensional convection},
  author = {Vieweg, P. P. and Schneide, C. and Padberg-Gehle, K. and Schumacher, J.},
  journal = {Phys. Rev. Fluids},
  volume = {6},
  issue = {4},
  pages = {L041501},
  numpages = {9},
  year = {2021},
  month = {Apr},
  publisher = {American Physical Society},
  doi = {10.1103/PhysRevFluids.6.L041501}
}

@article{KPG_RBC2022,
    author = {Schneide, C. and Vieweg, P. P. and Schumacher, J. and Padberg-Gehle, K.},
    title = {Evolutionary clustering of {L}agrangian trajectories in turbulent {R}ayleigh–{B}énard convection flows},
    journal = {Chaos: An Interdisciplinary Journal of Nonlinear Science},
    volume = {32},
    number = {1},
    pages = {013123},
    year = {2022},
    month = {01},
    issn = {1054-1500},
    doi = {10.1063/5.0076035}
}

@article{Vera2019,
  title = {Coherent {L}agrangian swirls among submesoscale motions},
  author = {Beron-Vera, F.J. and Hadjighasem, A. and Xia, Q. and Olascoaga, M.J. and Haller, G.},
  journal = {Proceedings of the National Academy of Sciences of the United States of America},
  volume = {116},
  issue = {37},
  pages = {18251-18256},
  year = {2019},
  month = {Sep},
  doi = {10.1073/pnas.1701392115},
}

@book{acheson1990elementary,
  title     = {Elementary Fluid Dynamics},
  author    = {Acheson, D. J.},
  isbn      = {9780198596790},
  lccn      = {89022947},
  year      = {1990},
  publisher = {Clarendon Press},
  edition   = {1st},
}

@article{VIEWEG202469,
title = {Lagrangian studies of coherent sets and heat transport in constant heat flux-driven turbulent {R}ayleigh–{B}énard convection},
journal = {European Journal of Mechanics - B/Fluids},
volume = {103},
pages = {69-85},
year = {2024},
issn = {0997-7546},
doi = {https://doi.org/10.1016/j.euromechflu.2023.08.007},
author = {P. P. Vieweg and A. Klünker and J. Schumacher and K. Padberg-Gehle}
}

@article{Shiskina2010,
title = "Boundary layers structure in turbulent thermal convection and its consequences for the required numerical resolution",
author = "O. Shishkina and Stevens, R. {J.A.M.} and S. Grossmann and D. Lohse",
note = "Open access article",
year = "2010",
doi = "10.1088/1367-2630/12/7/075022",
language = "English",
volume = "12",
journal = "New journal of physics",
issn = "1367-2630",
publisher = "Institute of Physics",
number = "7",
}

@article{Gastine_Wicht_Aurnou_2015,
title={Turbulent {R}ayleigh–{B}énard convection in spherical shells},
volume={778},
DOI={10.1017/jfm.2015.401},
journal={Journal of Fluid Mechanics},
author={Gastine, T. and Wicht, J. and Aurnou, J. M.},
year={2015},
pages={721–764}
}

@article{Breuer2004,
  title = {Effect of inertia in {R}ayleigh-{B}\'enard convection},
  author = {Breuer, M. and Wessling, S. and Schmalzl, J. and Hansen, U.},
  journal = {Phys. Rev. E},
  volume = {69},
  issue = {2},
  pages = {026302},
  numpages = {10},
  year = {2004},
  month = {Feb},
  publisher = {American Physical Society},
  doi = {10.1103/PhysRevE.69.026302}
}

@article{GROSSMANN_LOHSE_2000,
title={Scaling in thermal convection: a unifying theory},
volume={407},
DOI={10.1017/S0022112099007545},
journal={Journal of Fluid Mechanics},
author={Grossmann, S. and Lohse, D.},
year={2000},
pages={27–56}
}

@article{AHLERS_2006,
title={Non-{O}berbeck–{B}oussinesq effects in strongly turbulent {R}ayleigh–{B}énard convection},
volume={569},
DOI={10.1017/S0022112006002916},
journal={Journal of Fluid Mechanics},
author={Ahlers, G. and Brown, E. and Araujo, F. F. and Funfschilling, D. and Grossmann, S. and Lohse, D.},
year={2006},
pages={409–445}
}

@article{Tilgner_1996,
  title = {High-{R}ayleigh-number convection in spherical shells},
  author = {Tilgner, A.},
  journal = {Phys. Rev. E},
  volume = {53},
  issue = {5},
  pages = {4847--4851},
  numpages = {0},
  year = {1996},
  month = {May},
  publisher = {American Physical Society},
  doi = {10.1103/PhysRevE.53.4847}
}

@article{karrasch2015,
    author = {Karrasch, D.},
    title = {Attracting {Lagrangian} coherent structures on {R}iemannian manifolds},
    journal = {Chaos: An Interdisciplinary Journal of Nonlinear Science},
    volume = {25},
    number = {8},
    pages = {087411},
    year = {2015},
    month = {08},
    issn = {1054-1500},
    doi = {10.1063/1.4928451}
}

@article{RBC_review_Lohse,
  title = {Ultimate {R}ayleigh-{B}\'enard turbulence},
  author = {Lohse, D. and Shishkina, O.},
  journal = {Rev. Mod. Phys.},
  volume = {96},
  issue = {3},
  pages = {035001},
  numpages = {60},
  year = {2024},
  month = {Aug},
  publisher = {American Physical Society},
  doi = {10.1103/RevModPhys.96.035001},
}

@article{Vallis_Parker_Tobias_2019,
title={A simple system for moist convection: the {R}ainy–{B}énard model},
volume={862},
DOI={10.1017/jfm.2018.954},
journal={Journal of Fluid Mechanics},
author={Vallis, G. K. and Parker, D. J. and Tobias, S. M.},
year={2019},
pages={162–199}
}

@article{sun_convection_review,
   author = "Hanasoge, S. and Gizon, L. and Sreenivasan, K. R.",
   title = "Seismic Sounding of Convection in the Sun", 
   journal= "Annual Review of Fluid Mechanics",
   year = "2016",
   volume = "48",
   number = "Volume 48, 2016",
   pages = "191-217",
   doi = "https://doi.org/10.1146/annurev-fluid-122414-034534",
   publisher = "Annual Reviews",
   issn = "1545-4479",
   type = "Journal Article",
  }

@article{LSC_reversals,
  title = {Cessations and reversals of the large-scale circulation in turbulent thermal convection},
  author = {Xi, H.-D. and Xia, K.-Q.},
  journal = {Phys. Rev. E},
  volume = {75},
  issue = {6},
  pages = {066307},
  numpages = {5},
  year = {2007},
  month = {Jun},
  publisher = {American Physical Society},
  doi = {10.1103/PhysRevE.75.066307},
}

@article{haller_LCS_review,
   author = "Haller, G.",
   title = "Lagrangian Coherent Structures", 
   journal= "Annual Review of Fluid Mechanics",
   year = "2015",
   volume = "47",
   number = "Volume 47, 2015",
   pages = "137-162",
   doi = "https://doi.org/10.1146/annurev-fluid-010313-141322",
   url = "https://www.annualreviews.org/content/journals/10.1146/annurev-fluid-010313-141322",
   publisher = "Annual Reviews",
   issn = "1545-4479",
   type = "Journal Article",
  }

@article{Haller_Beron-Vera_2013,
title={Coherent {L}agrangian vortices: the black holes of turbulence},
volume={731},
DOI={10.1017/jfm.2013.391},
journal={Journal of Fluid Mechanics},
author={Haller, G. and Beron-Vera, F. J.},
year={2013},
pages={R4}
}

@article {Haller_vortices_Pacific,
      author = "R. Abernathey and G. Haller",
      title = "Transport by {L}agrangian Vortices in the {E}astern {P}acific",
      journal = "Journal of Physical Oceanography",
      year = "2018",
      publisher = "American Meteorological Society",
      address = "Boston MA, USA",
      volume = "48",
      number = "3",
      doi = "10.1175/JPO-D-17-0102.1",
      pages= "667 - 685"
}

@article{SHADDEN2005,
author = {S. C. Shadden and F. Lekien and J. E. Marsden},
title = {Definition and properties of {Lagrangian} coherent structures from finite-time {Lyapunov} exponents in two-dimensional aperiodic flows},
journal = {Physica D: Nonlinear Phenomena},
volume = {212},
number = {3},
pages = {271-304},
year = {2005},
issn = {0167-2789},
doi = {https://doi.org/10.1016/j.physd.2005.10.007}
}

@article{Rypina2017,
  title={Trajectory encounter volume as a diagnostic of mixing potential in fluid flows},
  author={Rypina, I. I. and Pratt, L. J.},
  journal={Nonlin. Processes Geophys.},
  volume={24},
  number={2},
  pages={189--202},
  year={2017},
  publisher={Copernicus GmbH}
}

@article{Badza2023,
  author  = {A. Badza and T. W. Mattner and S. Balasuriya},
  title   = {How sensitive are {Lagrangian} coherent structures to uncertainties in data?},
  journal = {Physica D: Nonlinear Phenomena},
  volume  = {444},
  pages   = {133580},
  year    = {2023},
  issn    = {0167-2789},
  doi     = {10.1016/j.physd.2022.133580}
}

@article{Allshouse2015,
  author  = {Allshouse, M. R. and Peacock, T.},
  title   = {Lagrangian based methods for coherent structure detection},
  journal = {Chaos},
  year    = {2015},
  volume  = {25},
  number  = {9},
  eid     = 097617,
  pages   = {097617},
  doi     = {10.1063/1.4922968}
}

@article{Hadjighasem2017,
  author  = {Hadjighasem, A.  and Farazmand, M. and Blazevski, D.  and Froyland, G.  and Haller, G.},
  title   = {A Critical Comparison of {Lagrangian} Methods for Coherent Structure Detection},
  journal = {Chaos},
  volume  = {27},
  number  = {5},
  pages   = {053104},
  year    = {2017}
}

@article{Balasuriya2018,
title = {Generalized {L}agrangian coherent structures},
journal = {Physica D: Nonlinear Phenomena},
volume = {372},
pages = {31-51},
year = {2018},
issn = {0167-2789},
doi = {https://doi.org/10.1016/j.physd.2018.01.011},
url = {https://www.sciencedirect.com/science/article/pii/S0167278917302750},
author = {S. Balasuriya and N. T. Ouellette and I. I. Rypina},
}

@article{curbelo2025,
  title={Lagrangian descriptors in geophysical flows: a survey},
  author={Curbelo, J.},
  journal={SeMA Journal: bolet{\'\i}n de la Sociedad Espa{\~n}ola de Matem{\'a}tica Aplicada},
  year={2025},
  publisher={Springer}
}

@article{Curbelo2023,
author = {Curbelo, J. and Rypina, I. I.},
title = {A Three Dimensional {L}agrangian Analysis of the Smoke Plume From the 2019/2020 {A}ustralian Wildfire Event},
journal = {Journal of Geophysical Research: Atmospheres},
volume = {128},
number = {21},
pages = {e2023JD039773},
doi = {https://doi.org/10.1029/2023JD039773},
year = {2023}
}

@article{rivoire2026,
  title={Tracking jet streams as {L}agrangian objects},
  author={Rivoire, L. and Curbelo, J. and Linz, M.},
  journal={Communications Earth \& Environment},
  volume={7},
  number={1},
  pages={267},
  year={2026},
  publisher={Nature Publishing Group UK London}
}

@article{mancho2013,
  title={Lagrangian descriptors: A method for revealing phase space structures of general time dependent dynamical systems},
  author={Mancho, A. M. and Wiggins, S. and Curbelo, J. and Mendoza, C.},
  journal={Communications in Nonlinear Science and Numerical Simulation},
  volume={18},
  number={12},
  pages={3530--3557},
  year={2013},
  publisher={Elsevier}
}

@article{bruera2023,
  title={Mixing and geometry in the {N}orth {A}tlantic meridional overturning circulation},
  author={Bruera, R. and Curbelo, J. and Garc{\'\i}a-S{\'a}nchez, G. and Mancho, A. M.},
  journal={Geophysical Research Letters},
  volume={50},
  number={7},
  pages={e2022GL102244},
  year={2023},
  publisher={Wiley Online Library}
}

@article{curbelo2021,
  title={Lagrangian analysis of the northern stratospheric polar vortex split in {A}pril 2020},
  author={Curbelo, J. and Chen, G. and Mechoso, C. R.},
  journal={Geophysical Research Letters},
  volume={48},
  number={16},
  pages={e2021GL093874},
  year={2021},
  publisher={Wiley Online Library}
}

@article{arantxa2003,
    author = {Net, M. and Alonso, A. and Sánchez, J.},
    title = {From stationary to complex time-dependent flows at moderate Rayleigh numbers in two-dimensional annular thermal convection},
    journal = {Physics of Fluids},
    volume = {15},
    number = {5},
    pages = {1314-1326},
    year = {2003},
    month = {05},
    issn = {1070-6631},
    doi = {10.1063/1.1565335}
}

@article{arantxa1998,
title = {Onset of convection in a rotating annulus with radial gravity and heating},
journal = {Fluid Dynamics Research},
volume = {24},
number = {3},
pages = {133-145},
year = {1999},
issn = {0169-5983},
doi = {10.1016/S0169-5983(98)00020-3},
author = {A. Alonso and M. Net and I. Mercader and E. Knobloch}
}

\end{document}